\documentclass[aps,prd, onecolumn,nofootinbib,10pt]{revtex4}
\usepackage{graphicx}
\usepackage{epsfig}
\usepackage{amsfonts}

\usepackage{amsmath}
\usepackage{amssymb}
\usepackage{mathrsfs}
\usepackage{graphicx}
\usepackage{float}
\usepackage{caption}

\usepackage{ulem}
\usepackage{color}

\newcounter{fig}   \newcommand{\lbfig}[1]{\refstepcounter{fig}
\label{#1} }
\newcommand{\vphi}{\varphi}

\newcommand{\bea}{\begin{eqnarray}}
\newcommand{\eea}{\end{eqnarray}}
\newcommand{\be}{\begin{equation}}
\newcommand{\ee}{\end{equation}}

\def\vecp{{\pmb{p}}}

\def\vecnabla{{\pmb{\nabla}}}

\newcommand{\re}[1]{(\ref{#1})}

\begin{document}

\title{Chains of interacting solitons}
\author{Ya.~Shnir}
\affiliation{BLTP, JINR, Dubna 141980, Moscow Region, Russia}

\begin{abstract}
We present an overview of multisoliton chains arising in various non-integrable field theories, and discuss
different mechanisms, which may lead to the occurrence of such axially-symmetric classical solutions. We explain
the pattern of interactions between different solitons, in particular Q-balls, Skyrmions and monopoles and show
how chains of interacting non-BPS solitons may form in a dynamic equilibrium between repulsive and attractive
forces.
\end{abstract}
\maketitle

%%%%%%%%%%%%%%%%%%%%%%%%%%%%%%%%%%%%%%%%%%%%%%%%%%%%%%%%%%
\section{Introduction}
%%%%%%%%%%%%%%%%%%%%%%%%%%%%%%%%%%%%%%%%%%%%%%%%%%%%%%%%%%
There has been astonishing progress over the past sixty years in our understanding of nonlinear phenomena.
Till 1960s nonlinear systems were given little attention mainly because of their complexity, in the most cases the
corresponding dynamical equations do not possess analytical solutions. The situation changed drastically with the
dawning of computational physics, which made it possible to find reasonably accurate solutions for nearly any
properly formulated physical problem. This development yields many surprising results and discoveries.

One of the most interesting properties of various non-linear systems is that they may support solitons, stable,
non-dissipative, localized configurations, behaving in many ways like particles (for review, see, e.g.,
\cite{Manton:2004tk,Babelon,Shnir2018}). Solitons emerge in diverse contexts in various situations in nonlinear optics,
condensed matter, nuclear physics, cosmology, and supersymmetric theories. In some situations, their existence is
related to topological properties of the model; in other cases they appear due to balance between the effects of
nonlinearity and dispersion. However, unlike the usual particles, solitons are extended objects, they possess a
core in which most energy is localized, and an asymptotic tale, which is responsible for the long-range interaction
between the well-separated solitons. Further, there is a tower of linearized
excitations around a soliton, which belong to the perturbative spectrum. As a result, the pattern of interaction
between the solitons becomes very complicated, in some situations there are both repulsive and attractive forces
with different asymptotic behavior and the process of the collision between the solitons is very different from
the simple picture of elastic scattering of point-like particles.

The study of the interactions between solitons, the processes of their scattering, radiation and annihilation
has attracted a lot of attention in many different contexts. First, almost immediately after discovery of the
solitons in pioneering works \cite{FPU,ZK}, the mathematical concept of integrability was developed. It turns out
that some models which may support solitons, are completely solvable, in other words, all solutions can be
presented analytically in closed form. Moreover, in integrable theories the collision between the solitons is
always completely elastic. Further, there is a very special class of so called self-dual solitons, whose
exactly saturate the topological energy bound. In such a case the energy of interaction between the solitons is
always zero, then various multisoliton configurations can be constructed via implication of diverse beautiful
differential-geometrical mathods, see e.g. \cite{Manton:2004tk,Babelon}.

The situation is completely different in non-integrable theories. Perhaps one of the simplest examples of such
theory is a family of (1+1)-dimensional models with a polynomial potential possessing two or more degenerated
minima, for example $\phi^4$ modes with a double-well potential \cite{Panos}. Another example is the Skyrme model
\cite{Skyrme:1961vq}, it is very well-known as a prototype of a relativistic field theory which supports
non-BPS topological solitons, see, for example \cite{Manton:2004tk}. Historically, the 3+1 dimensional Skyrme model was
proposed as a model of atomic nuclei, in such  a framework baryons are considered as solitons with identification
of the baryon number and the topological charge of the field configuration. A new development is related to the
lower dimensional version of the Skyrme model in 2+1 dimensions since solitons of that type were experimentally observed in planar
magnetic structures and liquid crystals (for a review, see e.q. \cite{sk}). Notably, a simplest Skyrmion of topological
degree one is rotationally invariant, however, solutions of higher degrees may possess very interesting geometric shapes
\cite{Manton:2004tk,Houghton:1997kg}. The reason  is that the interaction between the Skyrmions is mediated by the
long-range dipole forces, there are both repulsive and attractive channels. Thus, the pattern of interaction between the solitons
become rather involved. Further, a particular choice of the potential of
the model defines the structure of multi-soliton configurations, in some cases pairs of solitons in
equilibrium may appear.

Static solutions of the  $\phi^4$ theory and the Skyrme model represent class of topological solitons. There are
also non-topological solitons, for example stationary field configurations, commonly named Q-balls, that may
exist in some models  with a suitable self-interaction potential \cite{Rosen,Friedberg:1976me,Coleman:1985ki}.
When Q-balls are coupled to gravity so-called boson stars emerge, they represent compact stationary
configurations with a harmonic time dependence of the scalar field and unbrocken global symmetry
\cite{Kaup:1968zz,Ruffini:1969qy}. Notably, the character of the interaction between Q-balls depends on their relative
phase \cite{Battye:2000qj,Bowcock:2008dn}. In general, a pair of Q-balls is not stable in Minkowski spacetime,
however the gravitational attraction may stabilize it. Further, extended linear chains of rotating boson stars
can be formed via this mechanism \cite{Herdeiro:2020kvf}.

Another famous example of topological solitons are monopoles, they appear as classical solutions of the
non-abelian Yang-Mills-Higgs theory in 3+1 dimensions \cite{Hooft-Polyakov}, for a review see e.g. \cite{Smono}.
The magnetic charge of the non-abelian monopole is proportional to the topological charge. However, the pattern
of interaction between the monopoles is far from naive picture of Coulomb interaction of two point-like magnetic charges.
The 't Hooft-Polyakov solution is a coupled topologically stable configuration of gauge and Higgs fields
which may have different
asymptotic behavior. In particular, in the so-called Bogomol'nyi-Prasad-Sommerfield (BPS) limit of vanishing Higgs
potential, both the gauge and the scalar fields become massless, then there is an exact balance of two long-range
interactions between the BPS monopoles, which are mediated by the massless photon and the massless scalar
particle, respectively. In a contrary, attractive scalar interaction between two non-BPS monopoles becomes
stronger than magnetic repulsion, the corresponding charge two configuration with double zero of the Higgs field
at the origin possess axial symmetry \cite{Ward:1981jb}. On the other hand, it it possible to construct
monopole-antimonopole pair solution in a static equilibrium \cite{Rueber,MAP}. This configuration represent a
sphaleron, a saddle point solution of the classical field equations. Similar solutions also exist in the Skyrme
model \cite{Krusch:2004uf,Shnir:2009ct}.

Pairs of non-selfdual solitons in  a static equilibrium can be considered as basic building blocks of chains of
non-selfdual solitons. The main purpose of the present short review is to discuss such solution in a few
different models. First,  we briefly consider the  mechanism of interaction between the kinks in one spatial
dimension and describe how static multisoliton bound states can be formed due to exchange interaction mediated by
the localized fermion states (Section II). Then, we review the interactions between the Skyrmions and their
dependency on the structure of the potential. This yields some insight on the existence of chains of Skyrmions in
two and three spatial dimensions (Sections III and IV, respectively). In Section V we discuss interactions
between the Q-balls and possible mechanism of formation of pairs of Q-balls in equilibrium. Here we also include
the gravitational interaction and consider chains of boson stars. Finally, in Section VI, we revisit construction
of monopole-antimonopole pairs and chains. Conclusions and remarks are formulated in the last Section.

%%%%%%%%%%%%%%%%%%%%%%%%%%%%%%%%%%%%%%%%%%%%%%%%%%%%%%%%%%%%%%%%%%
\section{Chains of kinks}
%%%%%%%%%%%%%%%%%%%%%%%%%%%%%%%%%%%%%%%%%%%%%%%%%%%%%%%%%%%%%%%%%%
The simplest example of solitons are the kinks, they are classical solutions of the relativistic, nonlinear
scalar field theory in $1 + 1$ dimensions with Lagrangian density
\be
L=\frac12 \partial_\mu \partial^\mu \phi - V(\phi)
\label{lag-kinks}
\ee
with a smooth non-negative potential possessing some set of minima $V(\phi_0)=0$. The static
kink solution of this model interpolates between different vacua $\phi_0$ as space coordinate $x$
runs from $-\infty$ to $\infty$. For example, in the $\phi^4$ model the quartic potential
\be
V(\phi)= \frac12 \left(1-\phi^2\right)^2
\label{pot-phi4}
\ee
possesses two vacua, $\phi_0=\{\pm 1\}$. Then the field equation of the model \re{lag-kinks}
\be
\partial_\mu \partial^\mu \phi+\frac{\partial V}{\partial \phi}=0
\label{eqs-kink}
\ee
yields the well-known non-trivial kink solution $\phi_K(x)=\tanh x$ interpolating between $\phi(-\infty)=-1$
and $\phi(\infty)=1$. In a contrary, the anti-kink solution
$\phi_{\bar K}(x)=-\tanh x$ connects the vacua $\phi_0=1$ and $\phi_0=-1$.

There are other
examples of kinks in one spatial dimension, in particular the sine-Gordon model with infinitely degenerated
periodic potential $V(\phi)=1-\cos\phi$ supports static kink solution $\phi_K(x)=4 \arctan e^x$ which connects two
neighboring vacua
$\phi_0=0,\pi$. Similarly, the $\phi^6$ model with triple degenerated vacuum $V(\phi)=\frac12\phi^2\left(1-\phi^2 \right)^2$,
$\phi_0=\{0,\pm 1\}$, supports two different kinks
\be
\phi_{(0,1)}=\sqrt{\frac{1+\tanh x}{2}}\, ,\quad \phi_{(-1,0)}=-\sqrt{\frac{1-\tanh x}{2}}\,
\label{phi6-kinks}
\ee
Further examples of kinks are discussed in \cite{Manton:2018deu,Christov:2018ecz,Khare:2018psz} and other papers.

A few comments are in order here. First, for any model in one spatial dimension, the second order field equation
\re{eqs-kink} can always be reduced to the first order equation
\be
\frac{\partial \phi}{\partial x}=\pm \frac{\partial W}{\partial \phi}
\label{kink-bps}
\ee
where a {\it superpotential} $W(\phi)$ is defined as
$$
\frac12 \left(\frac{\partial W}{\partial \phi }\right)^2=V(\phi) \, .
$$
The equation \re{kink-bps}, often referred to as BPS equation,
is a simple realization of the self-duality of the model \re{lag-kinks}. In other words, kinks always saturate
the topological energy bound, their mass is proportional to the topological charge of the soliton.

Since our discussion focuses on chains of solitons, we are now looking for a possibility to construct
such static multisoltion solutions in one spatial dimensions. However, the energy of interaction between the kinks is not
zero, there is always a force acting
between the solitons. This force  can be evaluated when we consider an initial configuration of two widely
separated kinks, for example in the sine-Gordon model
\be
\label{initial-kink-SG}
\phi(x)=\phi_K(x+d)+\phi_K(x-d)-2\pi \, ,
\ee
where $\phi_K(x)$ is the kink solution mentioned above, and $d$ is the separation parameter.
Then we can
expand the corresponding energy of the configuration
in powers of $1/d$ and subtract the
mass of two infinitely separated kinks.
This yields the interaction energy
\be
\label{E_int}
E_{int} = 32 e^{-2d} \, .
\ee
Evidently, it is the Yukawa-type interaction, it is  repulsive for the kinks and it is attractive in the case of the
kink--anti-kink pair. Further, there is no multisoliton solution in the $\phi^4$ model with double degenerated vacuum,
the only possible structure of chains of solitons in such a case could be static linear configuration consisting of kinks and
anti-kinks in alternating order. However, the energy of interaction between the solitons is not zero again, by analogy with \re{E_int}
one can find that in the $\phi^4$ model $E_{int} = -16 e^{-4d}$, so there is an attractive force in the kink--anti-kink pair,
again. In other words, there is no static multisoliton solutions in the  model \re{lag-kinks}.

However, the situation can be different if we extend the model \re{lag-kinks}, for example we can
consider coupled two-component system with one of the scalar components having the kink structure and
the second component being a nontopological soliton \cite{Rajaraman:1978kd,Halavanau:2012dv}, or modify the model
\re{lag-kinks}, in such a way, that it still supports the kinks but possess a biharmonic  spatial derivative term
\cite{Decker:2020wnm}. In the latter case it is possible to construct a static kink-antikink pair. Similarly, such configuration
exist as a static solution of the  (1+1)  dimensional  scalar   field  theory coupled to  an  impurity \cite{Adam:2019djg} which
anchors the kinks.

Another way to construct a bounded kink-anti-kink pair is to include
interaction between the kinks and fermions \cite{Jackiw:1975fn,Chu:2007xh,Gani:2010pv,Klimashonok:2019iya,Perapechka:2019vqv}.
Such an  extended  model is defined by the following Lagrangian
\begin{equation}
\mathcal{L}=\frac{1}{2}\partial_\mu\phi\partial^\mu \phi
+ \bar \psi\left[ i\gamma^\mu \partial_\mu  -m - g\phi \right]\psi
-U(\phi) \, ,
\label{lag-ferm}
\end{equation}
where the self-interacting real scalar field $\phi$ is coupled with a
two-component Dirac spinor $\psi$ and $m,g$ are the bare
mass of the fermions and the Yukawa coupling constant, respectively. The matrices $\gamma_\mu$ are
$\gamma_0=\sigma_1$, $\gamma_1=i\sigma_3$ where $\sigma_i$ are the Pauli matrices, and
$\bar \psi = \psi^\dagger\gamma^0$.

Let us show that the coupling with fermions may provide an additional force stabilizing the kink--anti-kink pair.
We make use of the usual parametrization for a two-component spinor
$$
\label{ansFer}
\psi= e^{-i\epsilon t}\left(
\begin{array}{c}
u(x)\\
v(x)
\end{array}
\right)\, ,
$$
it results in the following coupled system of dynamical equations
\be
\begin{split}
\phi_{xx} + 2g uv -U^\prime &=0\, ;\\
u_x+(m+g\phi)u&=\epsilon v\, ;\\
-v_x+(m+g\phi)v&=\epsilon u\, .
\end{split}
\label{eq2}
\ee
This system is supplemented by the normalization condition
$\int\limits_{-\infty}^\infty\! dx\, (u^2+v^2)\!=\!1$ which we impose
as a constraint on the system \re{eq2}. Clearly, in the decoupled limit $g=0$,
the  model \re{lag-ferm} is reduced to the scalar model \re{lag-kinks} which supports the kinks.
\begin{figure}
 \begin{center}
\includegraphics[width=0.30\textwidth, trim={60, 20, 100, 60}, clip = true]{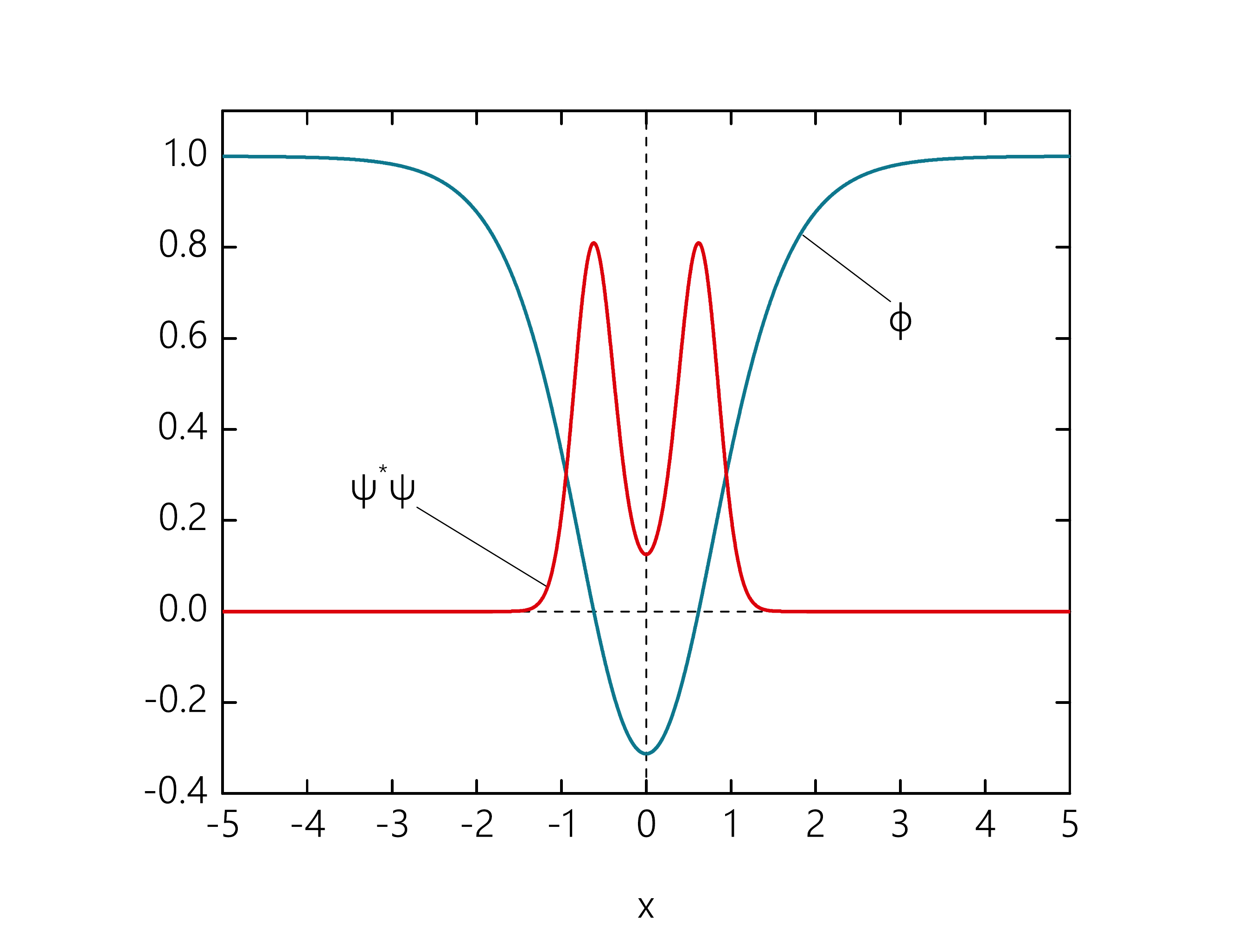}
\includegraphics[width=0.45\textwidth, trim={-40, -20, 0, 0}]{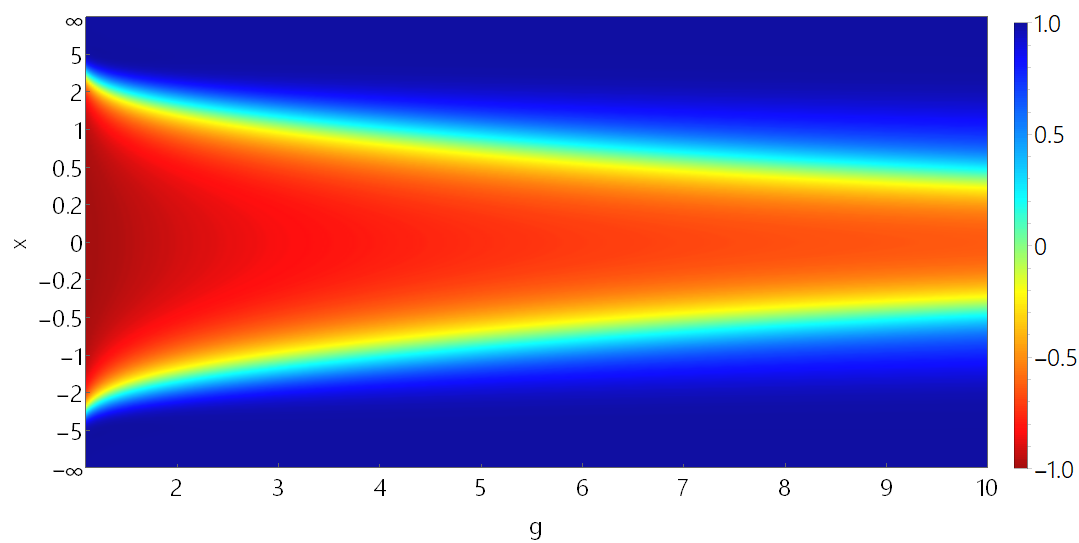}
\end{center}
 \caption{$\phi^4$ kink-antikink pair bounded by fermions. Profiles of the scalar field and
 fermion density distribution of the collective mode at $g=1$ (left plot) and
scalar field of the configuration bounded to this mode  vs Yukawa coupling $g$
(right plot). Reprinted (without modification) from \cite{Perapechka:2019vqv}, with permission of
APS.}
\lbfig{Fig1}
\end{figure}
We can easily see that for all such solutions, the system \re{eq2} possesses a
fermionic zero mode $\epsilon_0=0$ which is exponentially localized
on the kink. This mode exist for any value of the Yukawa coupling $g$,
there is no level crossing spectral flow in one spatial dimension \cite{Jackiw:1975fn}.
Notably,
for large values of the Yukawa coupling,
other localized fermionic states with non-zero energy eigenvalues $| \varepsilon | < |g-m| $ may appear
in the spectrum \cite{Jackiw:1975fn,Chu:2007xh,Klimashonok:2019iya,Liu:2008pi}.

Consideration of the fermion modes bounded to a kink usually
is related with an assumption that the back-reaction of the localized fermions is negligible
\cite{Jackiw:1975fn,Chu:2007xh}.
However, coupling to the higher localized modes may significantly distort the $\phi^4$ kink
\cite{Klimashonok:2019iya}. Further,
since such exponentially localized fermion modes may occur in multisoliton systems, localized
fermions could mediate the exchange interaction between the solitons.

Indeed, numerical solution of the full system of dynamical equations \re{eq2} shows that,
as the Yukawa coupling increases slightly above zero, a non-topological soliton emerge in the scalar sector,
this lump is linked to a localized fermionic mode extracted from the positive continuum \cite{Perapechka:2019vqv}.
As $g$ increases further, the lump becomes larger, it represents tightly bounded kink--anti-kink pair,
as seen in Fig.~\ref{Fig1}.

Further, we found collective
fermions localized on various multi-kink configurations. For example, a tower of localized fermion modes exist on a coupled
pair of sG kinks in the sector of topological degree two \cite{Perapechka:2019vqv}, similarly, there are bounded
pairs of  $\phi^6$ kinks and anti-kinks, see Fig.~\ref{Fig2}.

\begin{figure}
 \begin{center}
\includegraphics[width=0.35\textwidth, trim={49, 20, 100, 60}, clip = true]{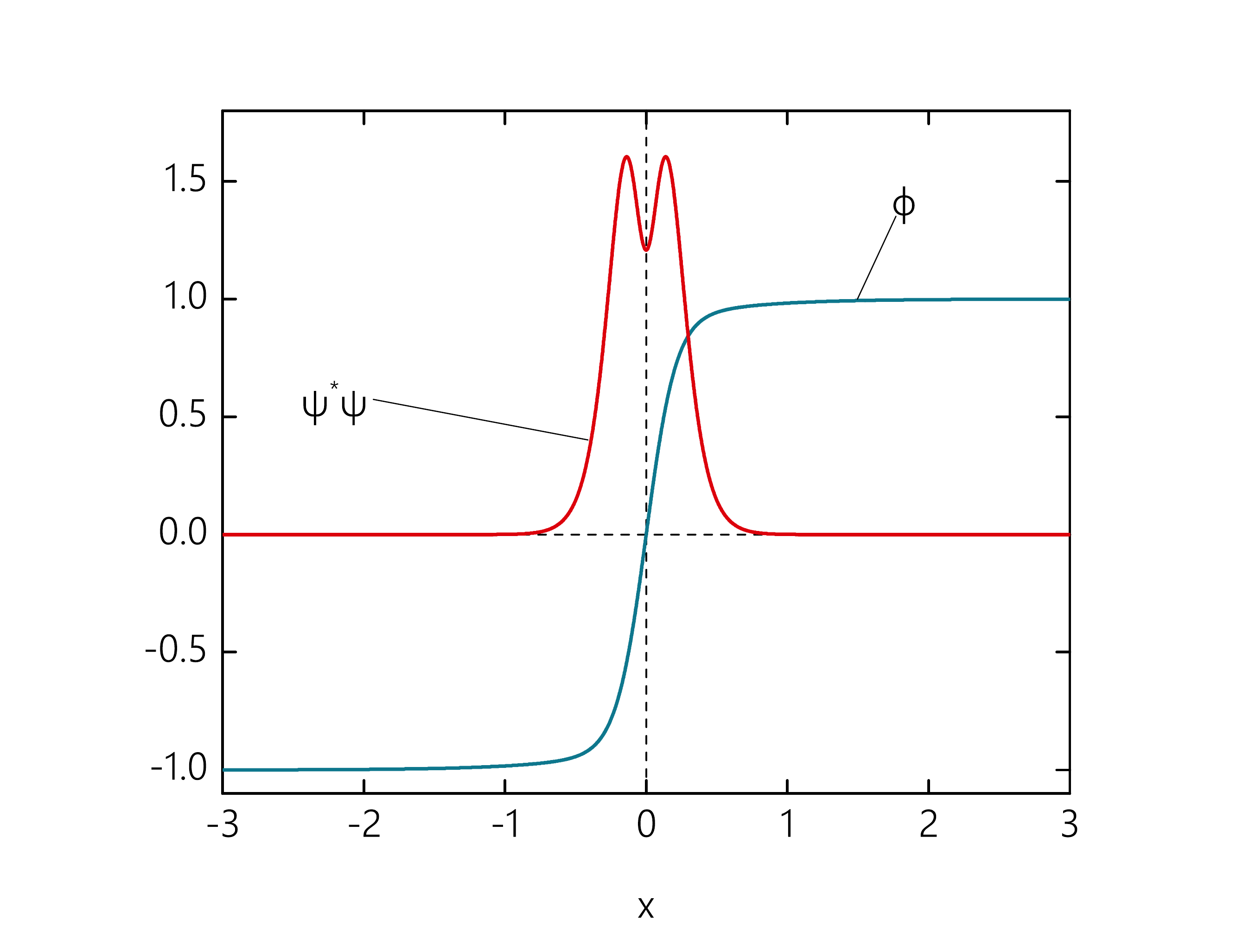}
\includegraphics[width=0.35\textwidth, trim={49, 20, 100, 60}, clip = true]{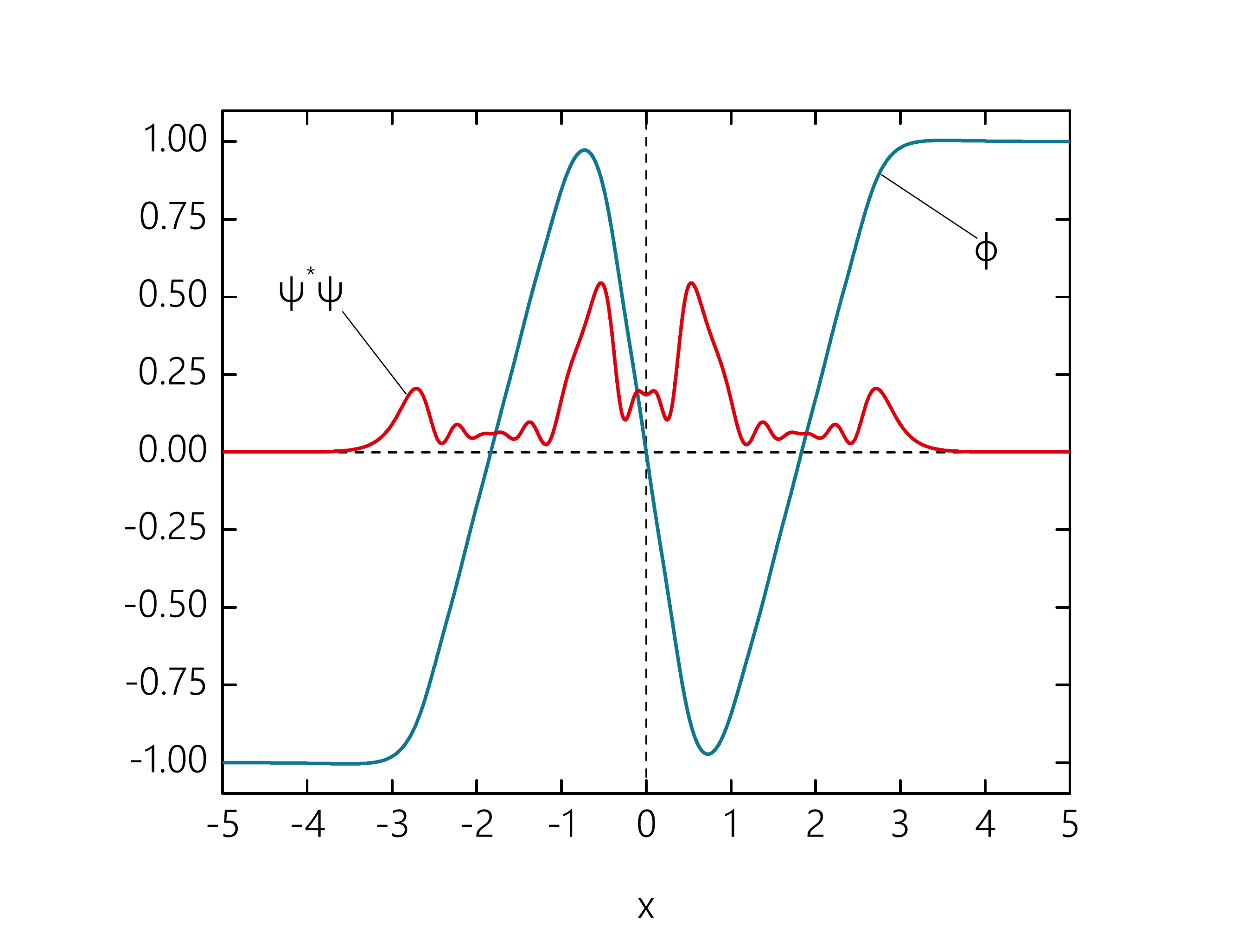}
\end{center}
 \caption{$\phi^6$ multikink configurations bounded by fermions. Profiles of the scalar field and
 fermion density distribution of the collective fermionic mode(left plot) and the chain of the
 kinks $(-1,1)+(1,-1)+(-1,1)$ bounded to the higher fermionic mode (right plot).  Reprinted (without modification) from \cite{Perapechka:2019vqv}, with permission of
APS.}
\lbfig{Fig2}
\end{figure}

Further, even more complicated bounded multisoliton configuration, which
represent multicomponent kink-antikink chains with localized fermion modes may exist in the extended model \re{lag-ferm}.
As a particular example, in the right plot of Fig.~\ref{Fig2}, we represent the chain of the
$\phi^6$ kinks $(-1,1)+(1,-1)+(-1,1)$ bounded by the higher fermion mode. Note that similar phenomena are
observed in other related models, the mechanism of the fermionic exchange interaction in multi-soliton congurations
is universal \cite{Perapechka:2019vqv}.

%%%%%%%%%%%%%%%%%%%%%%%%%%%%%%%%%%%%%%%%%%%%%%%%%%%%%%%%%%%%%%%%%%
\section{Chains of baby Skyrmions}
%%%%%%%%%%%%%%%%%%%%%%%%%%%%%%%%%%%%%%%%%%%%%%%%%%%%%%%%%%%%%%%%%%
Clearly, in one spatial dimension any bounded multisoliton configuration represents a chain of solitons. However, chains
of solitons exist in many other higher dimensional models. In all cases their existence is warranted due to a balance of
repulsive and attractive interactions between the solitons. In this section, as a simple example of such
configuration we will consider
2+1 dimensional planar Skyrme model \cite{Bogolubskaya:1989ha,Bogolyubskaya:1989fz,Leese:1989gj}. The model is defined by the
Lagrangian
\be
L=\frac12 (\partial_\mu \phi^a)^2 -
\frac{1}{4}\left(\varepsilon_{abc}\phi^a\partial_\mu \phi^b \partial_\nu \phi^c\right)^2
 - U(|\phi|) \, .
\label{lag-bsk}
\ee
Here the triplet of real scalar fields $\phi^a$, $a=1,2,3$  is constrained to the surface of unit sphere,
$\phi^a \cdot \phi^a =1$. In other words,
this is a topological map $\phi: S^2 \to S^2$ which is classified by the homotopy group $\pi_2(S^2)=\mathbb{Z}$.
The planar Skyrme model supports soliton solutions, which are classified in terms of the topological invariant:
\be
\label{winding-bsk}
Q = \frac{1}{8\pi} \int d^2x ~\varepsilon_{abc} \varepsilon_{ij}\phi^a \partial_i \phi^b \partial_j \phi^c \, .
\ee

The explicit choice of the potential term of the model \re{lag-bsk} is important because it defines the asymptotic form of the
field of the localized soliton. The most common choice is the $O(3)$ symmetry breaking potential
\be
\label{pot-old}
 U = \mu^2 (1-\phi_3) \, ,
\ee
where $\mu$ is the rescaled mass parameter. Indeed, the soliton of topological degree one can be constructed using the
rotationally invariant ansatz
\be
\label{rot-ansatz}
\phi_1 = \cos \theta \sin f(r);~~
\phi_2 = \sin \theta \sin f(r);~~
\phi_3 = \cos f(r)\, ,
\ee
%%%%%%%%%%%%%%%%%%%%%%%%%%%%%%%%%%%%%%%%%%%%%%%%%%%%%%%%%%%%%%%%%%%%%%%
\begin{figure}
\begin{center}
\setlength{\unitlength}{0.1cm}
%%\hspace{-5.2 cm}
\includegraphics[height=.22\textheight, angle =-0]{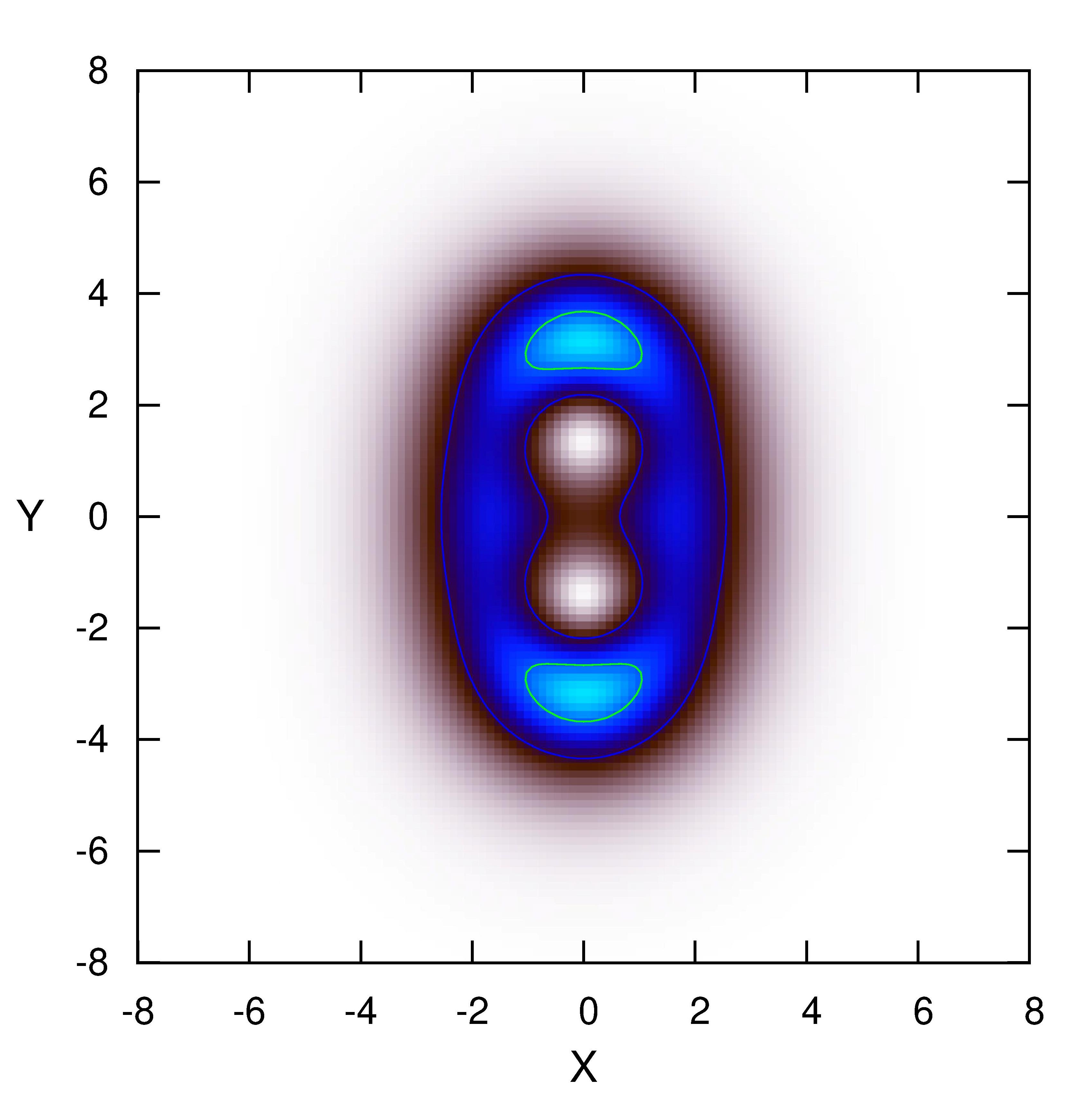}
\includegraphics[height=.23\textheight, angle =-0]{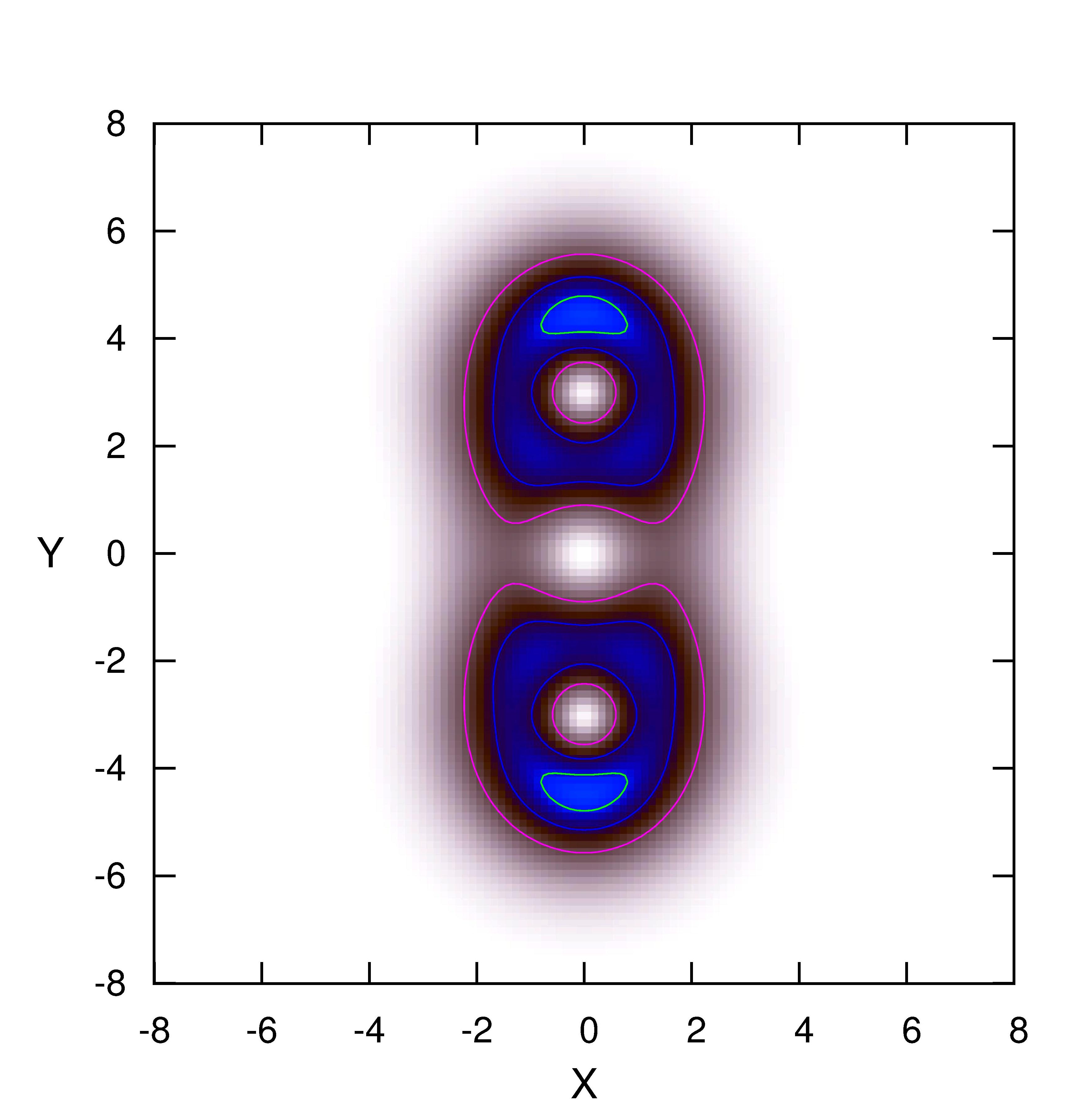}
\includegraphics[height=.24\textheight, angle =-0]{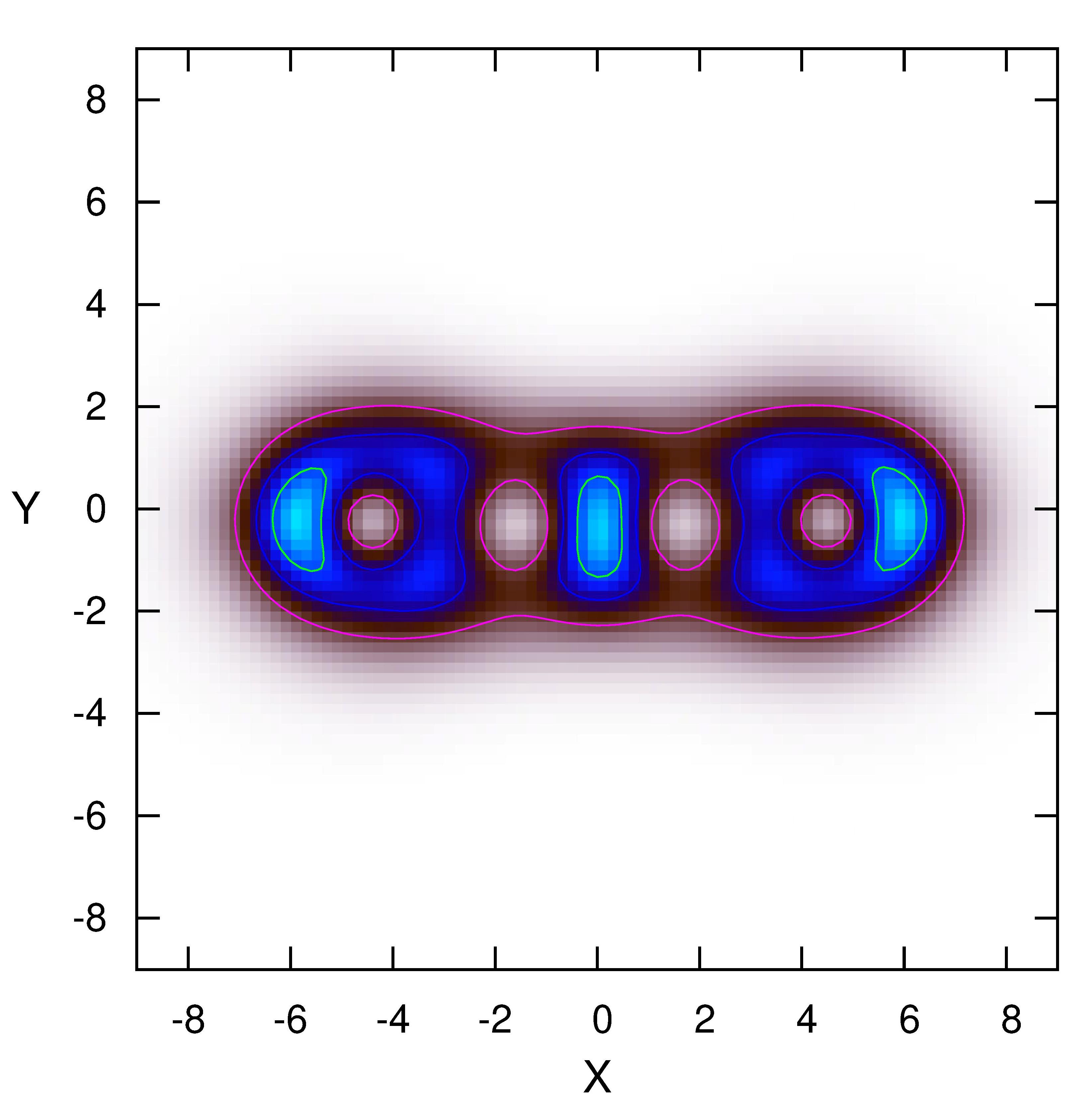}
\includegraphics[height=.14\textheight, angle =-0]{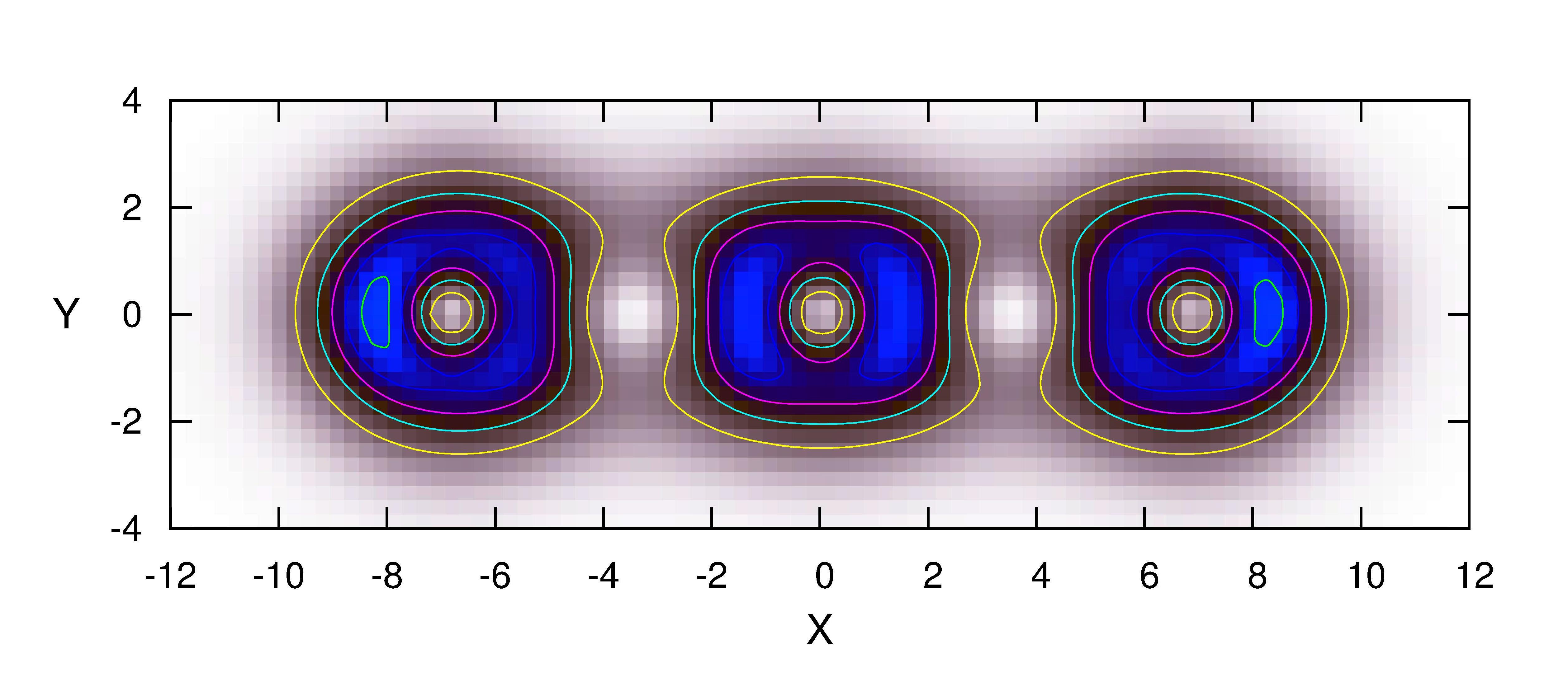}
\includegraphics[height=.14\textheight, angle =-0]{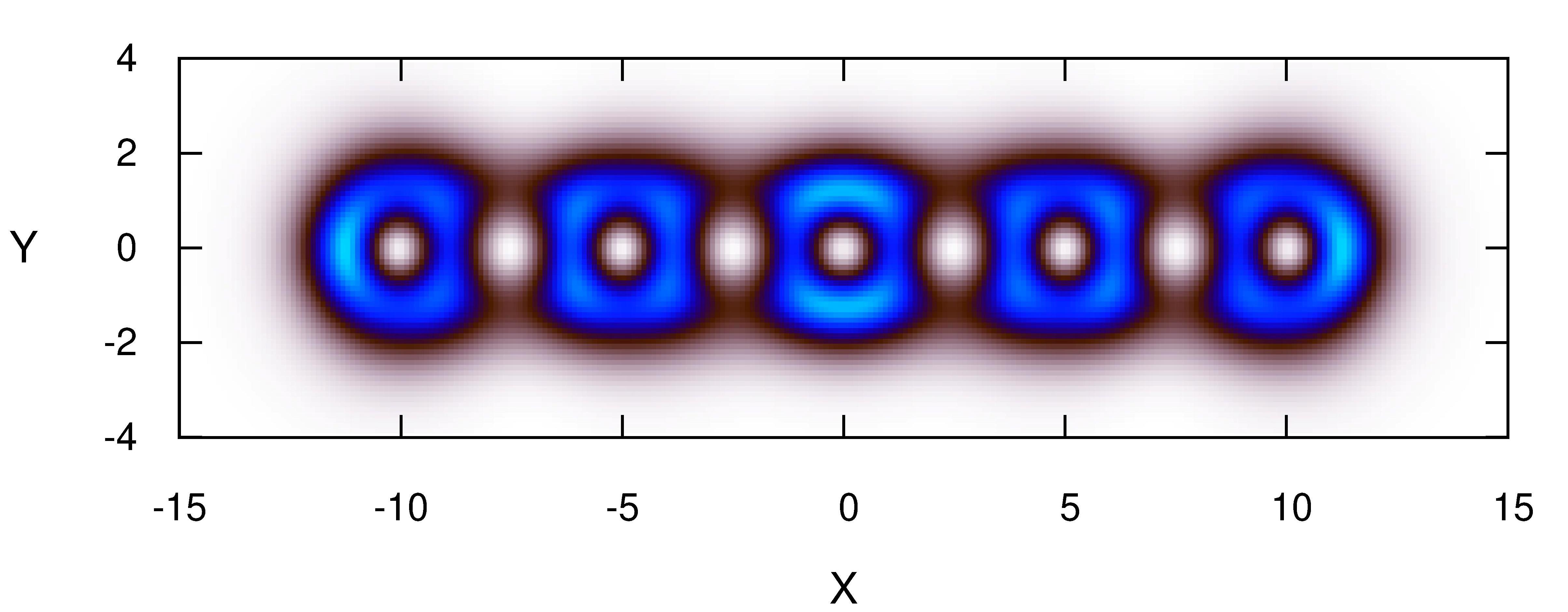}
\end{center}
 \caption{Contour plots of the energy density distributions of the solutions of the planar Skyrme model with the
potential \re{pot-old}
in the sectors of degrees $Q=3-6$ and $Q=10$.}
\lbfig{Fig3}
\end{figure}
where $f(r)$ is some monotonically decreasing
profile function. Since the field must approach the vacuum on the spacial
asymptotic, it satisfies the boundary condition $\cos f(r) \to 1$ as $r \to \infty$, i.e., $f(\infty) \to 0$. The system of field
equations of the baby Skyrme model \re{lag-bsk}
then is reduced to a single ordinary differential equation on the function $f(r)$:
\be
\label{eq-reduced}
\left(r + \frac{\sin^2 f}{r} \right)f^{\prime\prime} + \left(1- \frac{\sin^2 f}{r^2} + \frac{f^\prime \sin f
\cos f}{r}\right)f^{\prime}
- \frac{\sin f \cos f}{r} - r \mu^2 \sin f = 0 \, .
\ee

Linearizing this equation on the spatial infinity yields the asymptotic tail of the profile function
$f(r) \sim e^{-\mu r}/\sqrt r$. Evidently, this corresponds to the Yukawa-type decay with the
mass of scalar excitation $\mu$. On the other hand, the parameter $\mu$ defines the characteristic size of a Skyrmion, for
the potential \re{pot-old} the usual choice $\mu^2 = 0.1$ \cite{PZS} corresponds to the
localization of the energy within a region of diameter $r_0 \sim 1$. More precisely, the asymptotic
equation on the scalar field of the baby Skyrmion has the form
\be
\label{asymp-egs}
(\Delta - \mu^2 )\phi^a = \vecp^a \cdot  \vecnabla \delta(r)\, ,
\ee
Hence, the asymptotic field $\phi_a$ may be thought of as generated by a doublet of orthogonal dipoles, with
the strength $p$, one for each of the massive components  $\phi_1$ and $\phi_2$. The component $\phi_3$ remains
massless \cite{PZS}.

%%%%%%%%%%%%%%%%%%%%%%%%%%%%%%%%%%%%%%%%%%%%%%%%%%%%%%%%%%%%%%%%%%%%%
\begin{figure}
\begin{center}
\setlength{\unitlength}{0.1cm}
%%\hspace{-5.2 cm}
\includegraphics[height=.192\textheight, angle =-0]{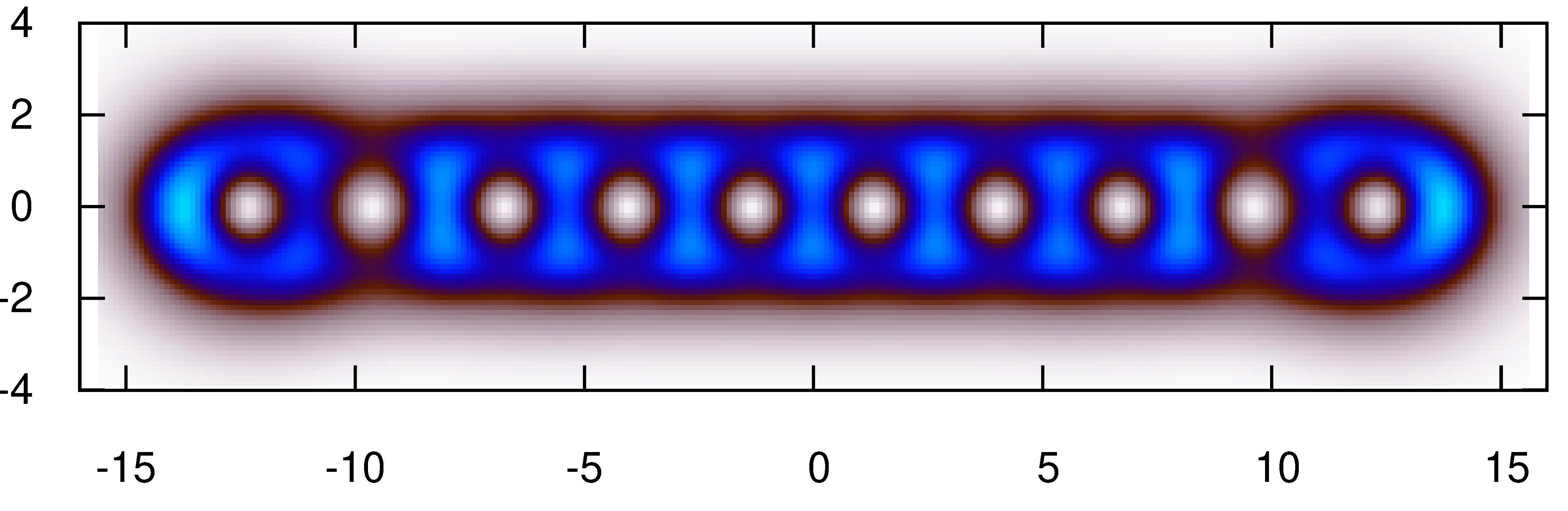}
\includegraphics[height=.18\textheight, angle =-0]{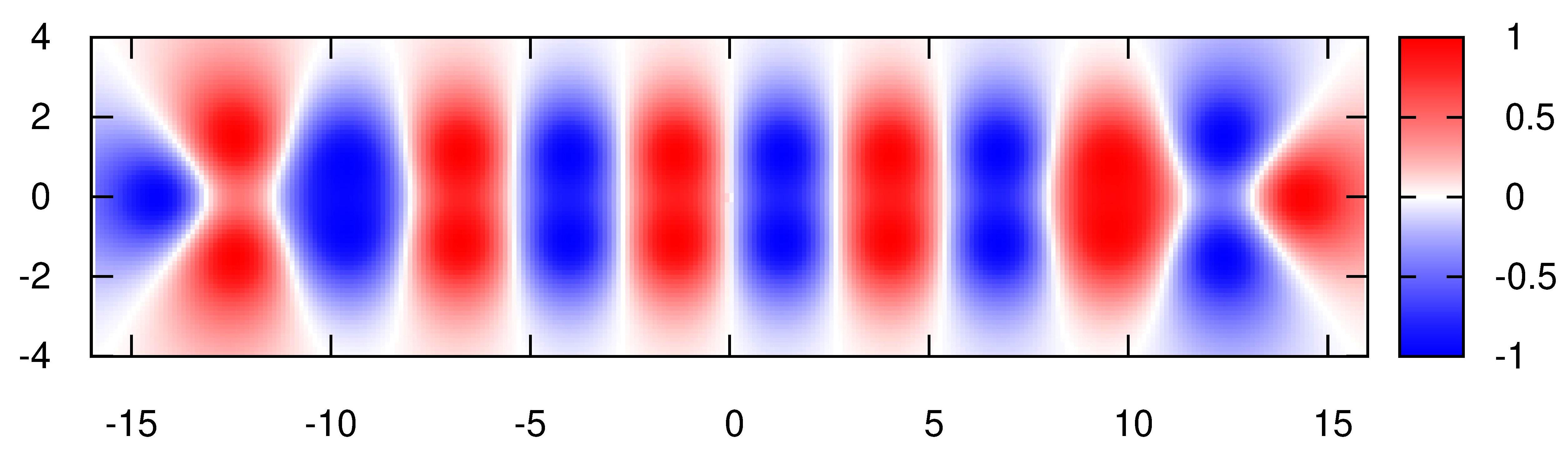}
\includegraphics[height=.18\textheight, angle =-0]{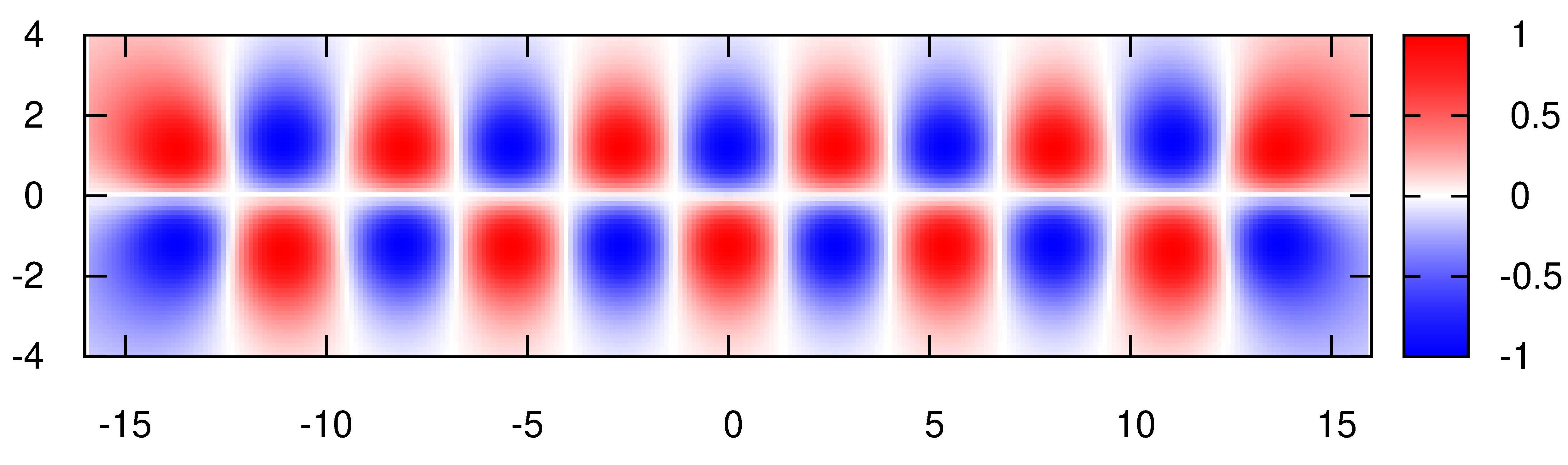}
\includegraphics[height=.18\textheight, angle =-0]{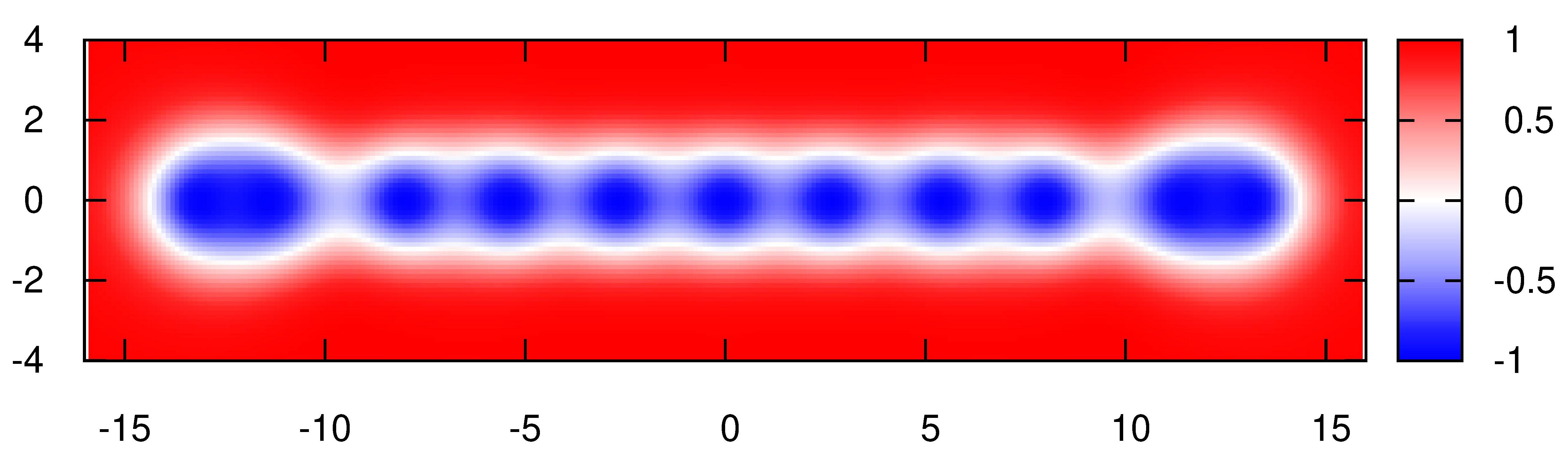}
\end{center}
 \caption{$Q=11$ chain of baby Skyrmions in the model \re{lag-bsk} with the potential \re{pot-old}:
Contour plots of the energy density distribution and the components of the field $\phi_1$, $\phi_2$ and $\phi_2$,
respectively, from top to bottom.}
\lbfig{Fig10}
\end{figure}

Let us now consider two widely separated unit charge Skyrmions. The leading term in the
energy of interaction of the solitons, evaluated by analogy with \re{E_int},
is
\be
E_{int}  ~\sim
\frac{\mu^2 p^2}{2\pi} \cos \chi \frac{e^{-\mu d}}{\sqrt d} \, ,
\ee
where the separation $d$ between the solitons is supposed to be much larger than the size of the core of the Skyrmion,
$d \gg 1/\mu$, and
$\chi$ is the relative angle of orientation of two pairs of orthogonal dipoles of equal strength $p$.

Thus, we conclude that the interaction of two separated Skyrmions is most attractive  when the two solitons are
exactly out of phase, $\chi=\pi$, and it is most repulsive when the relative phase $\chi=0$. However, nonlinear effects may
seriously affect this result even at intermediate separation of the solitons, one can expect some deformations of the core of the
solitons may induce a repulsive quasi-elastic force which may balance the long-range attraction.

Indeed, numerical simulations confirm, the model with the standard potential \re{pot-old} supports
existence of multi-soliton configurations \cite{PZS,Bsk,Weidig:1998ii}. First, as $\mu^2=0.1$,
the attractive force between two Skyrmions
of unit charge is stronger than repulsive quasi-elastic force, which is induced by deformations of the overlapping cores
of the solitons. Thus, the global minimum of the sector of degree two is also rotationally invariant, further, the dipole moment
of the $Q=2$ baby Skyrmion is zero.

Situation becomes different for baby Skyrmions of higher topological charges $Q\ge 3$
\cite{PZS,Bsk,Weidig:1998ii,Foster:2009vk}.
As $\mu^2=0.1$, the multisoliton solutions represent a stable chain of aligned, charge two baby skyrmions, see Fig.~\ref{Fig3}.
Another linear configuration of baby Skyrmions was constructed by Foster \cite{Foster:2009vk},
it represents a chain on charge one Skyrmion with both ends capped by two $Q=2$ solitons,
see Fig.~\ref{Fig10}.
In such a chain each soliton is rotated by $\pi$ with respect to its neighbor around the axis of symmetry.
The physical picture here is somewhat similar to the mechanism of formation of chain of dipoles in
classical electrodynamics, the energy of the chain is minimal with respect to all other possible configurations.
Notably, the linear configuration is formed dynamically. The energy of such a chain, which yields a global minimum of energy in
a given topological sector is slightly lower than the energy of the chain of
aligned $Q=2$ Skyrmions, for example the $Q=10$ chain of five aligned
Skyrmions of degree two displayed in Fig.~\ref{Fig3} has energy about $1 \%$ higher than the
Foster's cupped chain \cite{Foster:2009vk}.

Another way to construct linear chains of baby Skyrmions is to consider the
model \re{lag-bsk}  on a cylinder $\mathbb{R}^1\times S^1$ imposing anti-periodic
boundary conditions \cite{Harland:2007pb,Foster:2009vk}
\be
\label{anti-period-bsk}
\left(\phi_1(x,y+\beta),~\phi_2(x,y+\beta),~\phi_3(x,y+\beta)\right)=
\left(-\phi_1(x,y),~-\phi_2(x,y),~\phi_3(x,y)\right)\, ,
\ee
where $\beta$ is the period of the chain. This parametrization fixes a relative phase $\chi=\pi$ between the neighboring
Skyrmions. Then the energy of an infinitely charged chain becomes
a function of the periodicity $\beta$, it can be minimized to find lowest energy configuration.
Numerical evaluations suggest that for $\mu=1$ it corresponds to the period $\beta_{min}\approx 0.76 \pi$
\cite{Harland:2007pb}.

Some comments are in order here. First, linear periodic chains of planar Skyrmions exist
because of the balance of a short-range repulsion and a long-range attraction between two single solitons.
The long-range attraction is mediated by the dipole forces, however, the asymptotic form of the scalar field of the
baby skyrmion and the character of interaction between them, strongly depends on the particular choice of the
potential term $V(\phi)$. For example, the model \re{lag-bsk} with the double-vacuum potential $V(\phi)=\mu^2(1-(\phi_3)^2)$
\cite{Weidig:1998ii} always support rotationally invariant multi-soliton solutions. In a contrary, the choice of the holomorphic
potential $V(\phi)=\mu^2(1-\phi_3)^4$ \cite{Leese:1989gj} invariably yields repulsive interaction,
whatever the separation and relative
orientation of the Skyrmions \cite{Sutcliffe:1991aua} and there are no multisoliton solutions in such a model.
More complicated form of the potential may induce weak attraction, it allows for existence of
various multi-soliton configuration, including Skyrmions chains \cite{Salmi:2014hsa}.

Secondly, the existence of chains of baby Skyrmions is an intrinsic property of the
model \re{lag-bsk}, it is not necessarily to  modify it by analogy with 1+1 dimensional scalar model \re{lag-kinks}.
On the other hand, coupling to other fields may significantly affect the character of interaction between the solitons,
in particular, presence of fermionic modes localized by the baby Skyrmion yields two additional pairs of asymptotic dipoles
\cite{Perapechka:2018yux}, then the pattern of interaction between the solitons becomes more involved. Similarly, the asymptotic
forces in the gauged baby Skyrme model include contribution from the magnetic flux associated with the soliton
\cite{Samoilenka:2015bsf}. Even more complicated the pattern of interactions between planar Skyrmions becomes in the
$U(1)$ gauged baby Skyrme model with Chern-Simons term \cite{Samoilenka:2017}. However, for some set of parameters of the model,
there are solutions, which represent linear chains of electrically charged solitions with associated magnetic fluxes \cite{Samoilenka:2017}.

Note that Skyrmion configurations in 2+1 dimensions have recently been subject of considerable interest since solitons
of that type were experimentally observed in magnetic structures \cite{sk}. A magnetic Skyrmion is a
stable vortex-like configuration that exist in a thin film of chiral magnets, or in nematic crystals.
However, the usual Skyrme term in the model \re{lag-bsk} is replaced by the Dzyaloshinskii-Moriya chiral interaction term
\cite{Dzyaloshinskii,Moriya}. Similarly, such solutions exist in various condensed matter systems, in particular,
in chiral nematic liquid crystals \cite{Yu}. Further, these topological excitations minimize the Oseen-Frank free energy
functional, which also includes surface terms, see e.q. \cite{cond}.

\begin{figure}[h!]
\begin{center}
\setlength{\unitlength}{0.1cm}
%%\hspace{-5.2 cm}
\includegraphics[height=.38\textheight, angle =-0]{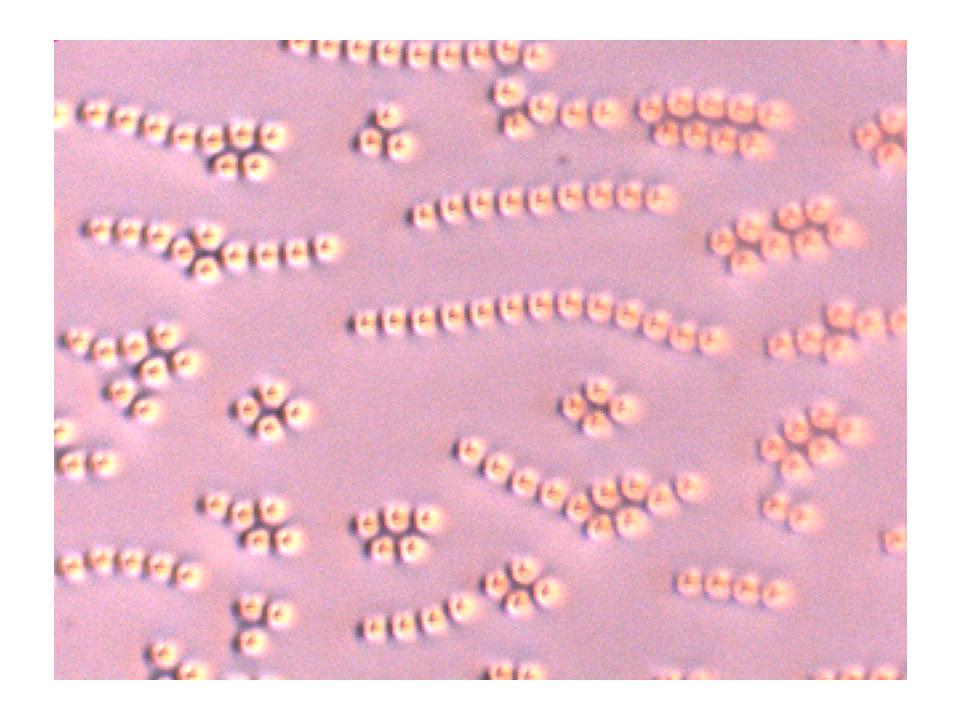}
\end{center}
 \caption{Chains of planar Skyrmions in a nematic crystal. (Courtesy of Ivan Smalyukh).}
\lbfig{Fig-Sk}
\end{figure}

Notably, there in no multisoliton solutions in the baby Skyrme model with  Dzyaloshinskii-Moriya term
supplemented by the Zeeman interaction term. The situation here is similar to the case of the
usual baby Skyrme model with holomorphic potential \cite{Leese:1989gj}, the interaction between chiral Skyrmions can be
only repulsive.
However, modification of the boundary conditions, or extension of the free energy functional allows for existence of
bounded multisoliton solutions \cite{Foster:2019rbd,Skmov,Schroers:2019hhe,Rybakov}.
In particular, a strong boundary electric field may generate chains of dynamical planar Skyrmions in a chiral nematic
crystal \cite{Skmov}, see Fig.~\ref{Fig-Sk}.

%%%%%%%%%%%%%%%%%%%%%%%%%%%%%%%%%%%%%%%%%%%%%%%%%%%%%%%%%%%%%%%%%%
\section{Chains of Skyrmions}
%%%%%%%%%%%%%%%%%%%%%%%%%%%%%%%%%%%%%%%%%%%%%%%%%%%%%%%%%%%%%%%%%%
The above-mentioned scheme of construction of baby Skyrmions can be extended to the original
Skyrme model in 3+1 dimensions.
The field of the model  is the unitary, unimodular matrix
$U({\bf r},t) \in SU(2)$, $U U^\dagger=\mathbb{I}$, which can be written
as an expansion in quartet of scalar fields $(\sigma, \pi^a)$ restricted to the surface of the sphere $S^3$:
\be
\label{Skyrme-field}
U=\sigma  + i \pi^a \cdot \tau^a ~~~ _{\overrightarrow{r\to \infty}}~~~  \mathbb{I} \, .
\ee
Here $\tau^a$ are the three usual Pauli matrices. Introducing the
quartet of scalar fields $\phi^a = (\sigma, \pi^1,\pi^2,\pi^3)$, restricted as
$\phi^a \cdot \phi^a =1$, we can write the Lagrangian of the model in the  form
\be
\label{lag-sk}
L=\partial_\mu \phi^a \partial^\mu \phi^a
-\frac12 (\partial_\mu \phi^a \partial^\mu \phi^a)^2 +
\frac12 (\partial_\mu \phi^a \partial_\nu \phi^a)(\partial^\mu \phi^b \partial^\nu \phi^b)
- \mu^2(1-\phi^a \phi^a_\infty) \, ,
\ee
where $\phi^a_\infty = (1,0,0,0)$. The vacuum boundary condition means that
the scalar field component $\sigma$ remains massless, while the triplet of pion fields $\phi_k$ has a mass $\mu$.
The topological charge of the Skyrmion is the winding number
\be
\label{Q-fields}
Q= -\frac{1}{12\pi^2}\int d^3x~ \varepsilon_{abcd}\varepsilon^{ijk}\phi^a \partial_i\phi^b\partial_j\phi^c\partial_k\phi^d \, .
\ee

Skyrmion solution of degree $Q=1$ is spherically symmetric, it can be constructed on the
hedgehog ansatz
\be
\label{hedg-ans}
U({\bf r})= e^{if(r) \hat r^a\cdot \tau^a } = \cos f(r) + i\sin f(r) \hat r^a\cdot \tau^a \, ,
\ee
where $f(r)$ is a real monotonically decreasing function of the radial variable with the boundary conditions
$f(0)=\pi$ and $f(\infty)=0$. Setting the boundary conditions
$f(0)=-\pi$ and $f(\infty)=0$ yields the winding number $Q=-1$, this is the anti-Skyrmion solution.
The profile function $f(r)$ satisfies the ordinary differential equation of second order
\be
\label{eq-f}
(r^2+2\sin^2f) f^{\prime\prime} + 2r f^\prime - \sin 2f \left(1-{f^\prime}^2 + \frac{\sin^2f}{r^2} \right)
+ \mu^2 \sin f = 0 \, .
\ee
The solution of this equation can be found numerically.

As it was outlined above, the character of long-range interaction between two separated solitons depends on the
asymptotic form of the field $\phi^a$. As $r\to \infty $,
$\cos f(r) \to 1$ and $\sin f(r) \sim f(r) \to 0$. Then the asymptotic form of the  solution of the equation
\re{eq-f} is
\be
f(r) \sim \frac{d}{4\pi  r^2} + ~O\left(\frac{1}{r^8}\right)\, , \qquad {\rm as}~ r\to \infty \, ,
\ee
where $d$ is some constant. Therefore the corresponding asymptotic triplet of massive pion fields, $\pi_i =
\sin f(r) \hat r_i$, represents the field of three mutually orthogonal scalar dipoles of equal dipole strength
$d$:
\be
\label{dip-3}
\pi_i = \frac{d r_i}{4\pi r^3} \, .
\ee
Consequently, the pattern of the long-distance interactions of two separated Skyrmions
depends on their relative orientation \cite{Schroers:1993yk}. If the solitons are aligned, they repel each other, by analogy
with above consideration of planar Skyrmions. The strongest repulsive force occurs as one of the Skyrmions
is rotated by $\pi$ about the axis joining the Skyrmions. The attractive channel in interaction of the solitons corresponds
to the case when one of the Skyrmions is rotated by $\pi$ about an axis perpendicular to axis $R$. However, for
the usual choice of the potential function in \re{lag-sk}, the attractive interaction in that channel is stronger than
quasi-elastic forces of deformation of the core, the resulting charge two configuration is axially-symmetric
\cite{Braaten:1988cc,Kopeliovich:1987bt,Manton:1987xf,Verbaarschot:1987au}.
Note that, asymptotically, the field of the axially symmetric $Q=2$ Skyrmion
has only one non-vanishing dipole component, associated
with the axis of symmetry of the configuration. Indeed, composing two Skyrmions into the axially
symmetric configuration, we cancel the dipole
fields which are orthogonal to the axis of symmetry while the components directed along this axis will add.
Thus, the dipole strength of the
$Q=2$ Skyrmion is approximately twice larger than $Q=1$ asymptotic dipole field.

In order to construct pair of bounded Skyrmions, one has to consider a nonstandard choice of potential term,
which combines both repulsive and attractive interactions \cite{Gillard:2015eia,Gillard:2016esy}. However, the dipoles forces
of the lightly bounded pair, which are orthogonal to the axis of symmetry,
do not support existence of a linear chain of Skyrmions of unit charge.

On the other hand, a Skyrme chain can be constructed in a way analogous to the construction of
arrays of baby Skyrmions outlined above \cite{Harland:2008eu}.
The field of the chain is supposed to be periodic in $z$-direction, i.e., $U(x,y,z)=R U(x,y,z+\beta)$ where matrix
of iso-rotations $R \in SO(3)$. Hence, each Skyrmion in the chain is iso-rotated by $R$ with respect to its neighbors.
Effectively, it corresponds to compactification of $z$-axis onto a circle $S^1$.
This construction can be thought of as a one-dimensional reduction of Skyrme crystals
\cite{Klebanov:1985qi,Kugler:1988mu,Castillejo:1989hq}.

Another possibility is to consider Skyrmion--anti-Skyrmion (SAS) pair in a static equilibrium, such configuration represents
a saddle point solution, a sphaleron \cite{Krusch:2004uf}. The SAS pair solution corresponds to a middle of
a non-contractible loop on the functional space of the Skyrme model. Physically, as said above,
the charge 2 axially-symmetric Skyrmion possesses only one asymptotic dipole field, the dipole interaction may stabilize the
SAS pair.

%%%%%%%%%%%%%%%%%%%%%%%%%%%%%%%%%%%%%%%%%%%%%%%%%%%%%%%%%%%%%%%%%%%%%%%
\begin{figure}
\begin{center}
\setlength{\unitlength}{0.1cm}
%%\hspace{-5.2 cm}
\includegraphics[height=.28\textheight, angle =-0]{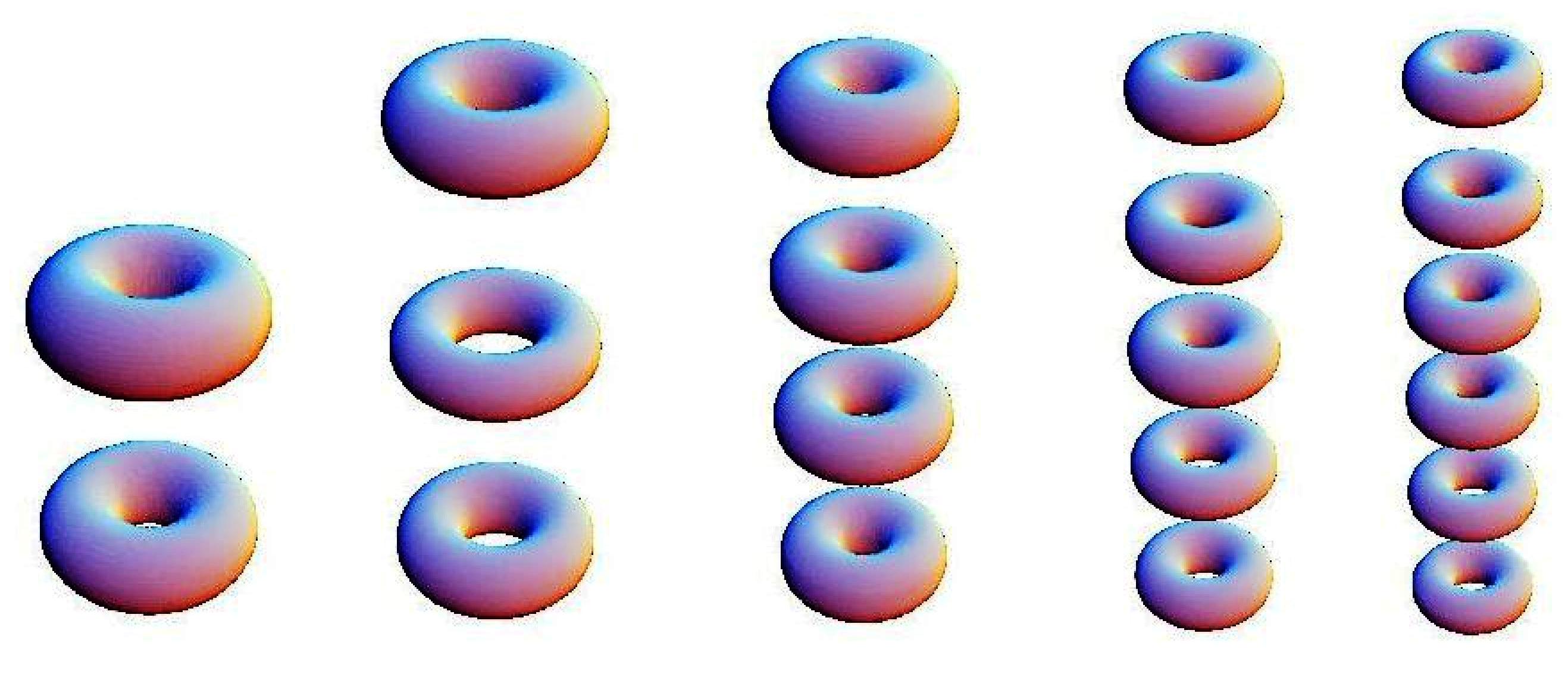}
\end{center}
\caption{Skyrmion--anti-Skyrmion chains. Reprinted (without modification) from \cite{Shnir:2009ct}, with STM Permission.}
\lbfig{Fig4}
\end{figure}

The SAS saddle point solution can be constructed numerically using the axially symmetric
parametrization of the field $U(\bf r)$
\be
\label{axial-sk}
U({\bf r}) = \cos f(r,\theta) + i \pi^a \cdot \tau^a \sin f(r,\theta) \, ,
\ee
where the triplet of pion fields is
\be
 \pi^1 = \sin g(r,\theta)\cos n \varphi;\quad
 \pi^2 = \sin g(r,\theta)\sin n \varphi;\quad
 \pi^3 = \cos g(r,\theta) \, ,
\label{pion-axial}
\ee
and $\sigma=\cos f(r,\theta)$. In this parametrization the integer $n \in \mathbb{Z}$ counts the
winding of the field in the $x-y$ plane.

For the SAS pair the boundary conditions imposed on the functions $f(r,\theta),~g(r,\theta)$ are
\be
\label{bc-Q2sas}
f(0,\theta)=\pi; \quad g(r,0)=0;\qquad f(\infty,\theta)=0;\quad g(r,\pi)=2 \pi \, ,
\ee
This yields a configuration with zero net topological charge, as one can see directly from \re{Q-fields}.
Further
generalization of this construction is possible, if we impose $g(r,\pi)=m \pi$, where $m$ is an integer number, which
counts the number of constituents in the resulting SAS chain configuration \cite{Shnir:2009ct}.
Together with the winding number $n$ of each individual Skyrmion appearing in \re{pion-axial}, it yields
the net topological charge of the axially symmetric chain:
\be
Q = \frac{n}{2}\left(1-(-1)^m\right) \, .
\ee
Clearly, the case $m = 1$ corresponds to the (multi)-Skyrmions
of topological charge $Q=n$, while $m = 2$ gives a pair with zero net topological charge
consisting of a charge $Q=n$ Skyrmion and a charge $Q=-n$ anti-Skyrmion. More general, for
odd values of $m$ the winding number $n$ coincides with the topological charge of the Skyrmion
$Q$ whereas even values of $m$ correspond to the deformations of the topologically trivial
sector. Thus we can construct a chain of charge $n$ Skyrmions
and charge $-n$ anti-Skyrmions placed along the axis of symmetry in alternating order.
Note that the solitons are dynamically arranged in such a linear configuration, in a contrary, the periodic Skyrmions
constructed on the ansatz $U(x,y,z)=R U(x,y,z+\beta)$, where $R$ is a rotation matrix,
by definition represent a one-dimensional cluster of equally
spaced Skyrmions. Such a chain, may only contract or extend itself  as the period $\beta$ varies \cite{Harland:2008eu}.

In Fig.~\ref{Fig4}  we represent the energy density isosurfaces of the $|Q|=2$ SAS chains
for zero pion mass. Notably,
there is no such a saddle point solution for a single $|Q|=1$  Skyrmion--anti-Skyrmion pair
\cite{Krusch:2004uf,Shnir:2009ct}. However, coupling to gravity may provide an additional attractive force, it stabilizes
the SAS pair in curved spacetime \cite{Shnir:2015aba,Shnir:2020hau}.

%%%%%%%%%%%%%%%%%%%%%%%%%%%%%%%%%%%%%%%%%%%%%%%%%%%%%%%%%%%%%%%%%%
\section{Chains of Q-balls and boson stars}
%%%%%%%%%%%%%%%%%%%%%%%%%%%%%%%%%%%%%%%%%%%%%%%%%%%%%%%%%%%%%%%%%%
In the previous sections we considered linear chains of topological solitons in various spatial dimensions.
Another type of chains can be constructed in models which support non-topological solitons.
One of the simplest examples in flat space
is given by Q-balls, stationary spinning configurations of a complex scalar field with a suitable
self-interaction potential \cite{Rosen,Friedberg:1976me,Coleman:1985ki}. When Q-balls are coupled to gravity
so-called boson stars emerge, which represent compact stationary configurations with a harmonic time dependence of the
scalar field \cite{Kaup:1968zz,Ruffini:1969qy}.

Both Q-balls and boson stars carry a Noether charge associated with an unbroken continuous global symmetry.
This charge is proportional to the angular frequency of the complex boson field and represents the
boson particle number of the configurations \cite{Friedberg:1976me,Coleman:1985ki}.

Localized Q-ball solutions appear in a simple 3+1 dimensional complex scalar model  with the
Lagrangian \cite{Coleman:1985ki}
\be
\label{lag-balls}
L= |\partial_\mu\phi|^2 - V(|\phi|) \, ,
\ee
and appropriate choice of the non-linear potential $V(|\phi|)$. The fundamental spherically symmetric
solution in this case represents a stationary spinning configuration $\phi= f({\bf r})e^{i\omega t}$ where
$f(r)$ is the real function of radial variable. This function satisfied the equation of motion
\be
\label{eq-radial}
\frac{d^2 f}{dr^2} + \frac{2}{r}\frac{df}{dr} + \omega^2 f =  \frac12 \frac{d U}{df} \, .
\ee
with the boundary conditions $\partial_r f(r)\left.\right|_{r=0}\!\!=\!0$ and $f(r)\left.\right|_{r=\infty}\!\!=\!0$.
Then the solution of the equation \re{eq-radial} must decay
asymptotically as
\be
f \sim \frac{1}{r} e^{-\sqrt{\mu^2-\omega^2} r} + ~O(1/r) \, ,
\ee
where $\mu^2 = \frac12 U^{\prime\prime}(0)$ is the mass of the scalar excitation.
In other words, the configuration is exponentially
localized at the origin.

Evidently, the properties of the Q-balls depend on the particular choice of the potential and
its parameters. It is convenient to make use of the non-linear sextic potential
\cite{Radu:2008pp,Kleihaus:2005me,Kleihaus:2007vk,Volkov:2002aj}
\be
\label{potential}
U(|\phi|)=a|\phi|^2 - b |\phi|^4 + c |\phi|^6 \, ,
\ee
where the positive parameters are taken as $a=1.1$, $b=2$ and $c=1$. Setting $b=c=0$ reduces the model to the usual
Klein-Gordon system in the flat space, which does not support any localized solutions.

%\begin{figure}[h!]
%    \begin{center}
%        \includegraphics[height=4.2cm]{BS_Q1-Q.jpg}
%        \includegraphics[height=4.2cm]{BS_Q1-Phi.jpg}
%        \includegraphics[height=4.2cm]{BS_Q2-Q.jpg}
%        \includegraphics[height=4.2cm]{BS_Q2-Phi.jpg}
%        \includegraphics[height=4.2cm]{BS_Q3-Q.jpg}
%        \includegraphics[height=4.2cm]{BS_Q3-Phi.jpg}
%        \includegraphics[height=4.2cm]{BS_Q4-Q.jpg}
%        \includegraphics[height=4.2cm]{BS_Q4-Phi.jpg}
%        \includegraphics[height=4.2cm]{BS_Q5-Q.jpg}
%        \includegraphics[height=4.2cm]{BS_Q5-Phi.jpg}
%    \end{center}
%    \caption{\small
%Chains of boson stars with one to five constituents (from top to bottom)
%on the fundamental branch for $\alpha=0.25$ at $\omega=0.80$:
%$3d$ plots of the $U(1)$ energy density distributions (left plots) and
%the scalar field functions $\phi$ (right plots)
%versus the coordinates $\rho = r\sin \theta$ and $z=r\cos\theta$.}
%    \lbfig{Fig5}
%\end{figure}

As the angular frequency tends to its upper critical value,  $\omega \sim 1$,
the Q-balls become linked to the perturbative excitations of the Klein-Gordon model,
$$
\phi \sim \frac{1}{\sqrt r} J_{l+\frac12}(r)Y_{ln}(\theta,\varphi)
$$
where $J_{l+\frac12}(r)$ is the Bessel function of the first kind of order $l$ and
$$
Y_{ln}(\theta,\varphi) =\sqrt{\frac{2l+1}{4\pi}\frac{(l-n)!}{(l+n)!} }P_l^n(\cos \theta) e^{in\varphi}
$$
are the usual spherical harmonics with $n\in [-l,l]$. Here $P_l^n(\cos \theta)$ are the associated
Legendre functions. The spherically symmetric fundamental Q-ball corresponds
to the spherical harmonic $Y_{00}$ while the simplest non-spherical excitation corresponds to the harmonic $Y_{10}$,
and induces a pair of oscillating perturbations with opposite phases. Such excitations can be considered as droplets of bosonic
condensate, they may exist in various models, in particular  in a Bose-Einstein condensate  with dipole-dipole interaction
\cite{BCdroplet}.

It was pointed out that, similarly to the  case of dipole-dipole interactions between the Skyrmions,
the character of the interaction between Q-balls in Minkowski spacetime depends on their relative phase
\cite{Battye:2000qj,Bowcock:2008dn}.  If the Q-balls are in phase, the interaction is attractive, if they are out of phase,
there is a repulsive force between them. Thus, an axially symmetric excitation of the complex scalar field of
the form $Y_{10}$ is in general not stable in Minkowski spacetime, however the gravitational interaction may stabilize it.
Consequently, a decrease of the angular frequency increases the size of the configuration producing a binary
system of boson stars spinning in opposite phases. Such a pair represents a building block of linear chains of Q-ball in
curved space-time \cite{BS2021,BSChains_2021}.

Let us now consider a self-interacting complex scalar field $\phi$, which is minimally
coupled to Einstein gravity. The corresponding action of
the system is
\be
\label{action}
S = \int{\sqrt{-g}\left(\frac{R}{4 \alpha^2}-L\right) d^4
x},
\ee
where $R$ is the Ricci scalar curvature with respect to the metric $g_{\mu\nu}$, $g$ denotes the
determinant of the metric, $\alpha^2=4\pi G$ is the gravitational coupling constant, $G$ is Newton's constant, and
$L$ is the matter field Lagrangian \re{lag-balls}. Below we consider axially-symmetric configurations, which can be
parameterized by the ansatz $\phi=\phi(r,\theta)e^{i\omega t}$ and make use of the Lewis-Papapetrou metric
\be
\label{metrans}
ds^2=-f dt^2 +\frac{m}{f}\left(dr^2+r^2 d\theta^2\right)+ r^2\sin^2 \theta \frac{l}{f}
d\varphi^2\,
\ee
where the  metric functions $f, m$ and $l$ are functions of $r$ and $\theta$ only.

Below we will consider axially symmetric configurations, composed of several constituents,
whose centers are localized on the symmetry axis. Such solutions can be obtained numerically \cite{BSChains_2021,BS2021}, the results
show that, indeed, there are multi-soliton linear solutions with $k$ nodes on the symmetry axis. For relatively small
values of gravitational coupling $\alpha$ and the angular frequencies a bit lower than the mass threshold,
all solitons assembled into a chain possess similar sizes, shapes and distance from their next neighbors.
However, this very democratic picture changes as we move along the set of
branches that form as the angular frequency $\omega$ is varied
(for a given coupling $\alpha$). In  Fig.\ref{Fig9} we displayed a few examples of
such chains of boson stars at $\alpha=0.25$ and $\omega=0.80$.

\begin{figure}
    \begin{center}
        \includegraphics[height=3.5cm]{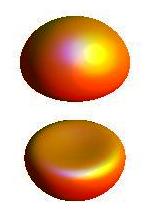}
        \includegraphics[height=4.5cm]{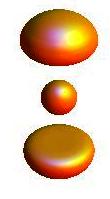}
        \includegraphics[height=5.5cm]{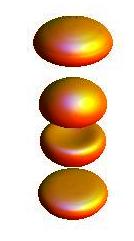}~~
        \includegraphics[height=6.5cm]{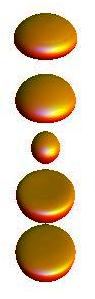}
    \end{center}
    \caption{\small
Chains of boson stars: Energy density isosurfaces
on the fundamental branch for $\alpha=0.25$ at $\omega=0.80$ (in different scales).}
    \lbfig{Fig9}
\end{figure}

As said above, pairs and chains of boson stars do not possess flat space limit, repulsive interaction between constituents of
the chain should be balanced by some attractive force. Such a force can appear if we consider spinning
$U(1)$ gauged Q-balls with non-zero angular momentum \cite{Loiko:2020htk}.

Notably, there are two families of the spinning Q-balls with positive and negative parity, the
corresponding solutions are symmetric or anti-symmetric with respect to reflections in the equatorial
plane \cite{Volkov:2002aj}. Apart the the above-mentioned fundamental spherically symmetric Q-balls,
there are both radially and angularly excited Q-balls  \cite{Volkov:2002aj,Kleihaus:2005me,Kleihaus:2007vk,Brihaye:2008cg}.
The radially excited solutions are still spherically symmetric, however
the scalar field possesses some set of radial nodes.
Such radially excited gauged Q-balls also exist in the $U(1)$ gauged model \cite{Loginov:2020lwg}. The angularly excited
solutions with some set of nodes in $\theta$-direction, can be parity-even, or parity-odd.

The angularly excited axially symmetric Q-balls with non-zero angular
momentum possess an additional azimuthal phase factor of the spinning field
\cite{Volkov:2002aj,Kleihaus:2005me,Radu:2008pp}. In the $U(1)$ gauged theory such configurations induce a toroidal
magnetic field \cite{Shiromizu:1998eh,Loiko:2019gwk}, these solutions can be viewed as vortons, the finite energy
localized spinning loops with non-zero angular momentum and magnetic flux
\cite{Witten:1984eb,Radu:2008pp,Davis:1988ij,Garaud:2013iba}.

We consider two-component $U(1)$ gauged Friedberg-Lee-Sirlin-Maxwell model \cite{Loiko:2019gwk,Loiko:2020htk},
which describes a coupled system of a real self-interacting scalar field $\psi$ and a
complex scalar field $\phi$, minimally interacting
with the Abelian gauge field $A_\mu$. The corresponding Lagrangian density is
\be
L= -\frac14 F_{\mu\nu} F^{\mu\nu} + (\partial_\mu\psi)^2 + |D_\mu\phi|^2 - m^2 \psi^2|\phi|^2 - U(\psi),
\label{lag-fls}
\ee
where $D_\mu = \partial_\mu +igA_\mu$ denotes the covariant derivative.
Here $F_{\mu\nu}=\partial_\mu A_\nu-\partial_\nu A_\mu$ is the electromagnetic field strength tensor,
$g$ is the gauge coupling constant and $m$ is the scalar coupling constant.
The symmetry breaking quartic potential of the real scalar field $\psi$ is
$
U(\psi)= \mu^2 (1-\psi^2)^2 \, ,
$
Notably, the model \re{lag-fls} can be considered as a generalization of the
Abelian Higgs model,  in other words gauged Q-ball behaves like a
superconductor \cite{Lee:1988ag} with the field component $\psi$ playing a role of the order parameter.

The Lagrangian \re{lag-fls} is invariant under the local $U(1)$ gauge
transformations of the fields, the corresponding conserved Noether current is
\be
\label{Noether}
j_\mu = i(\phi D_\mu\phi^\ast-\phi^\ast D_\mu\phi) \, .
\ee
This current is a source in the Maxwell equation
\be
\partial^\mu F_{\mu\nu}= g j_\nu
\label{eq-em}
\ee
Two other dynamical equations correspond to the variations of the Lagrangian \re{lag-fls} with respect to the
fields $\psi$ and $\phi$, respectively:
\be
\label{scaleq}
\begin{split}
    \partial^\mu\partial_\mu \psi&=-m^2 \psi |\phi|^2+2 \mu^2 \psi \left(1-\psi^2\right),\\
    D^\mu D_\mu \phi&=-m^2 \psi^2 \phi \, ,
\end{split}
\ee

\begin{figure}
\includegraphics[height=4.4cm]{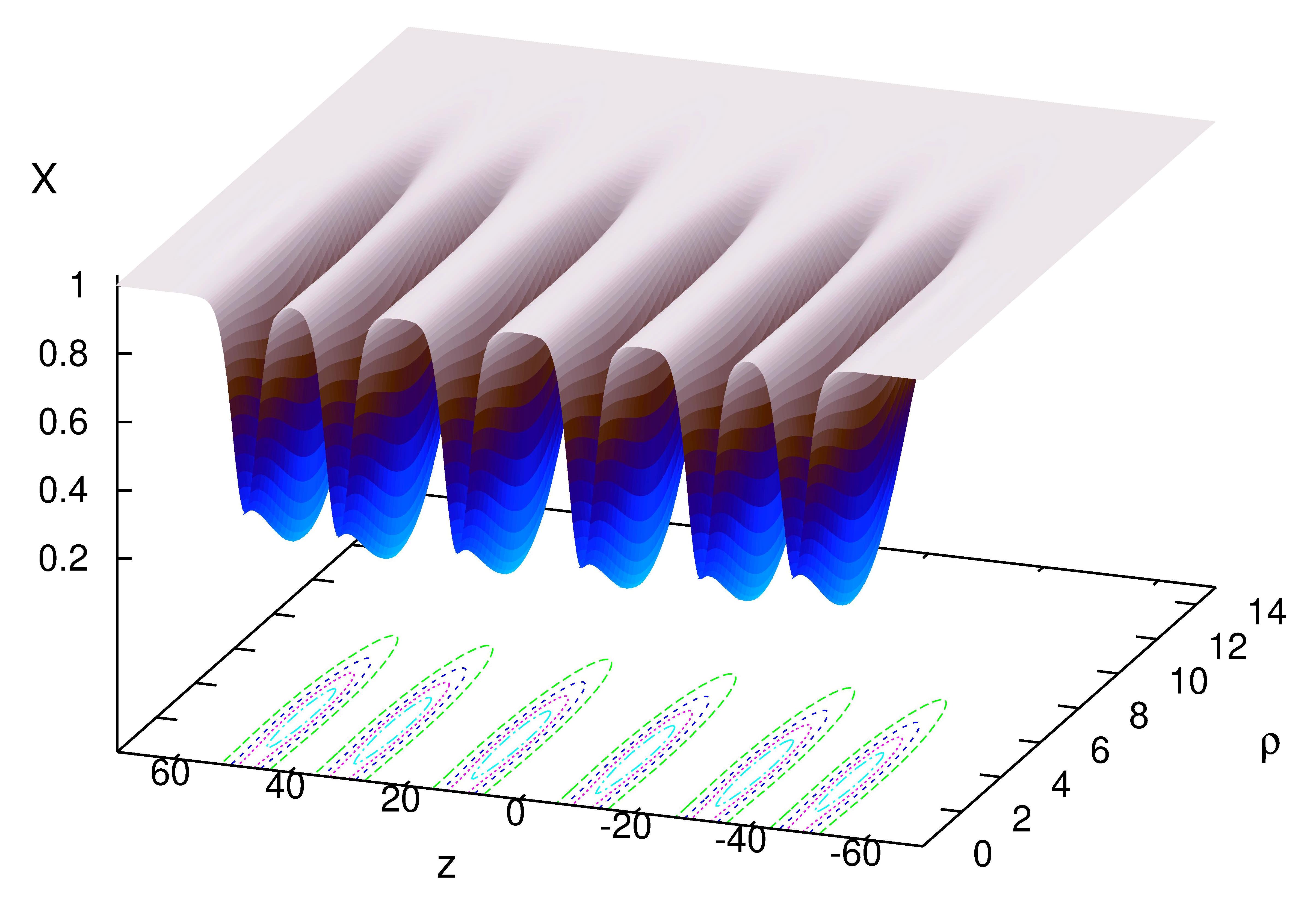}
\includegraphics[height=4.4cm]{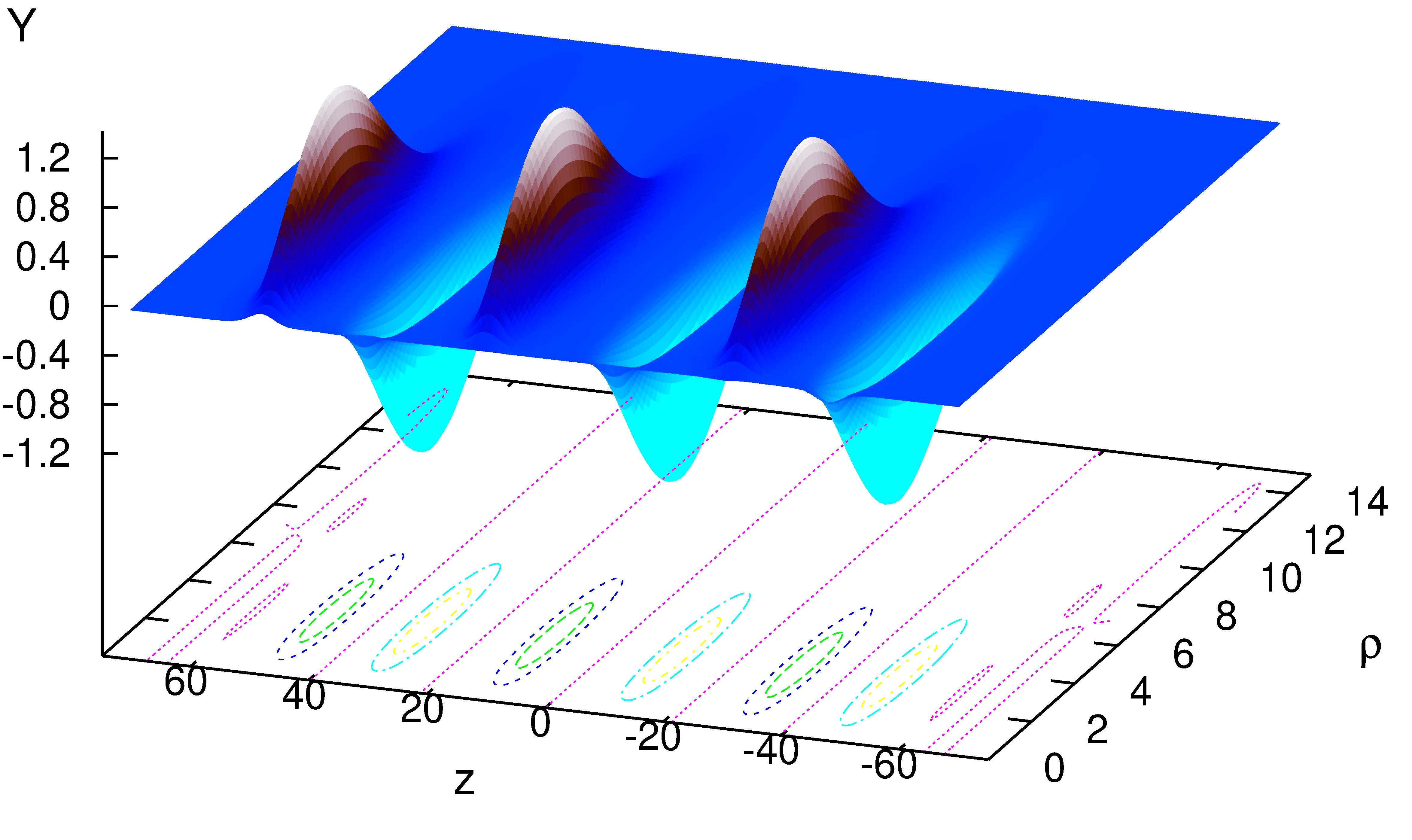}
\includegraphics[height=4.2cm]{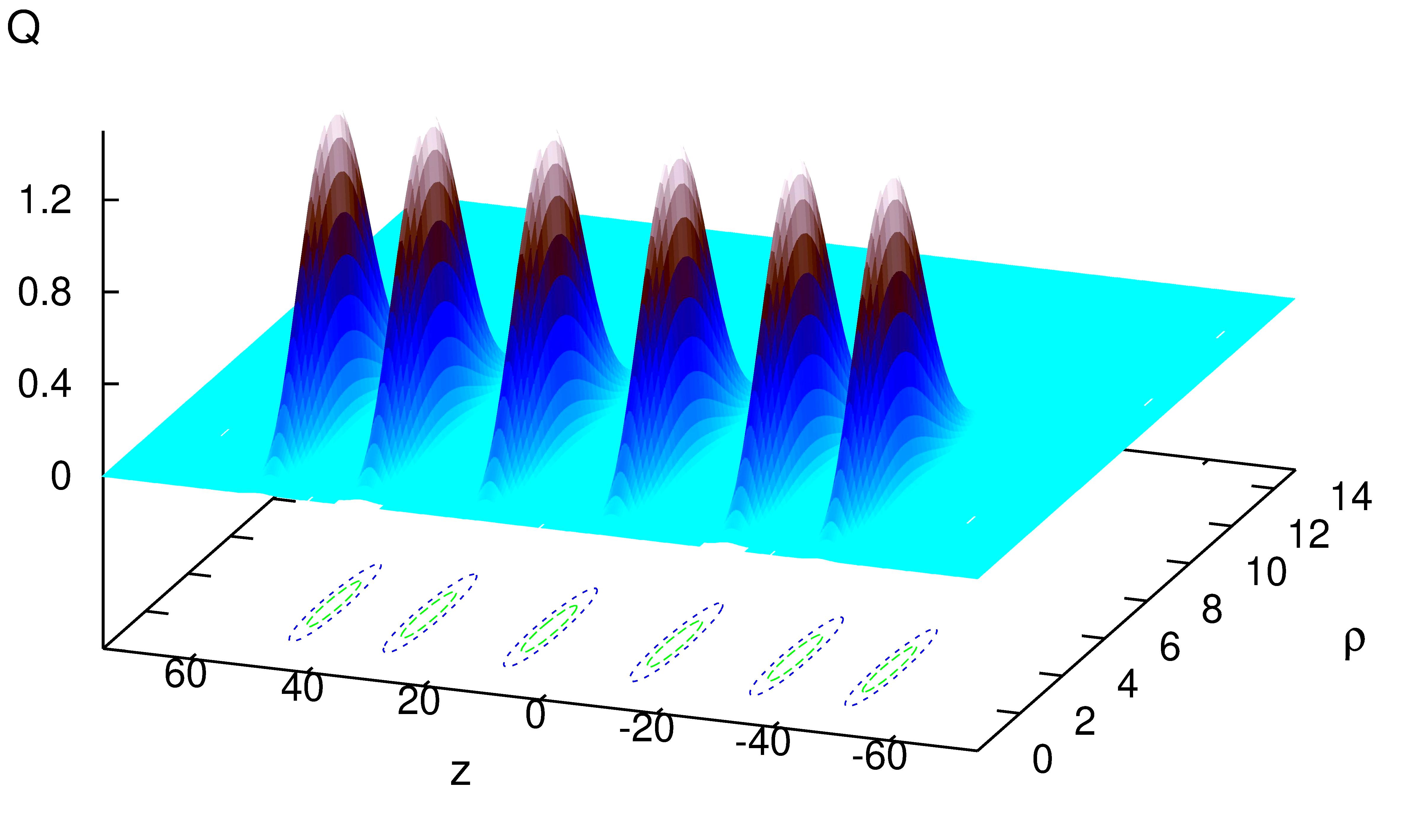}
\includegraphics[height=4.2cm]{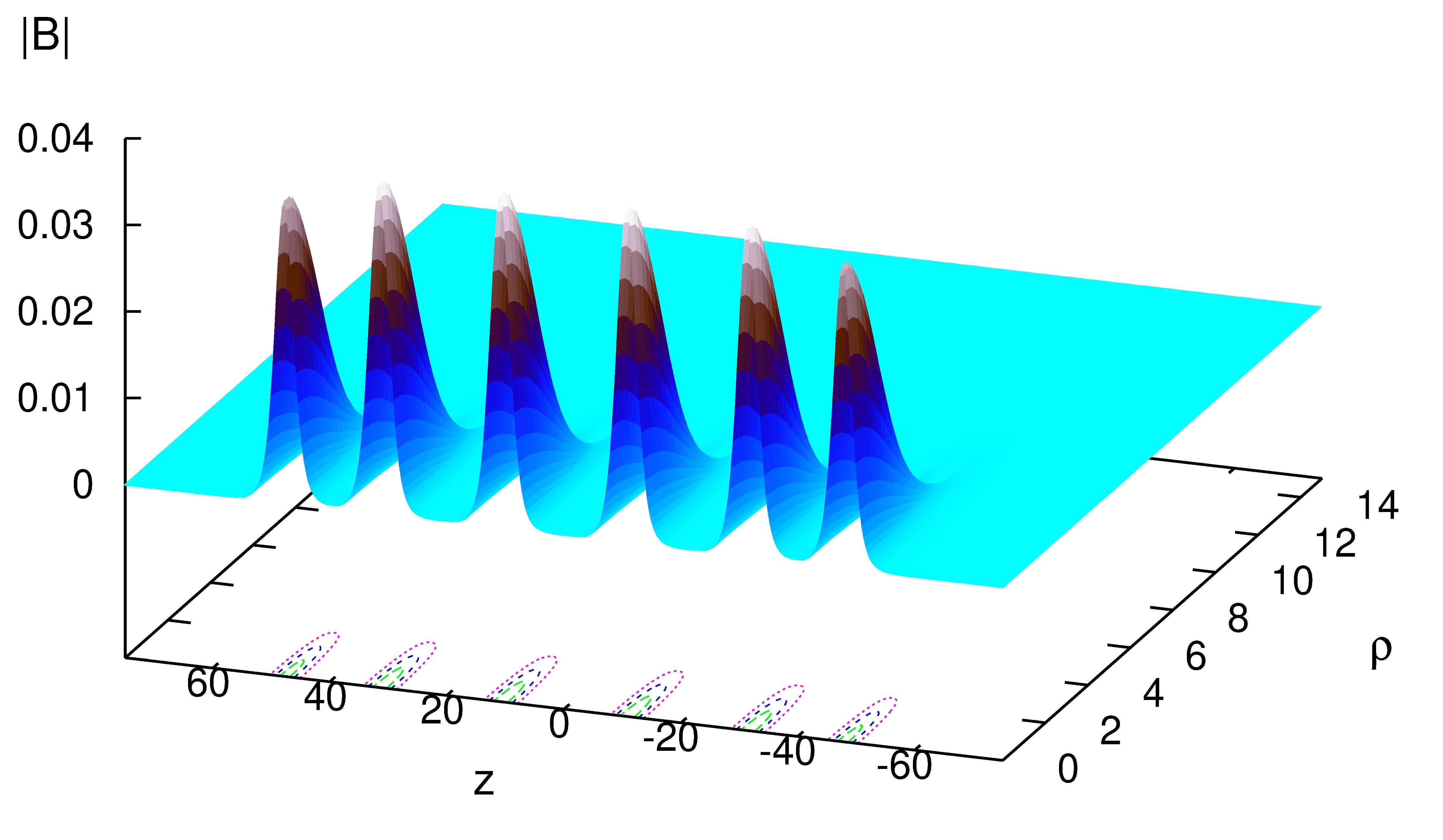}
\caption{\small
Axially-symmetric $n=1$ chain of gauged Q-balls:
The  field components  $X$ (upper left) and $Y$ (upper right),
the electric charge density distribution (bottom left),
and the magnitude of the magnetic field distribution (bottom right)
of the $k=6$ chain \cite{Loiko:2020htk}.}
\label{Fig6}
\end{figure}

Below we consider stationary spinning axially-symmetric solutions of the model \re{lag-fls}.
The corresponding  parametrization of the scalar fields is
\be
\label{scalans}
\psi=X(r,\theta)\, , \qquad  \phi=Y(r,\theta)e^{i(\omega t+n\varphi)}\, ,
\ee
where $\omega$ is the angular frequency of the spinning complex field $\phi$, and
$n\in\mathbb{Z}$ is the azimuthal winding number.
Further, in the static gauge the electromagnetic potential is
\be
\label{Aans}
A_{\mu} dx^{\mu} =A_0(r,\theta)dt + A_\varphi(r,\theta) \sin\theta d\varphi \, ,
\ee
hence, the spinning gauged Q-ball with non-zero angular momentum possess both electric charge and magnetic flux.
Clearly, both electric and magnetic field of the Q-ball contribute to the energy of interaction between two solitons,
together with Yukawa interactions mediated by the scalar fields.

In order to construct saddle point solution, which represent a pair of Q-balls in a static equilibrium, we have to balance the
attractive and repulsive forces. Note that the real scalar component $\psi$ may induce only attractive interaction,
while the spinning complex field $\phi$ of two Q-balls in opposite phases generates repulsive force. Further, electric
charge of the components of the pair always yields a repulsive Coulomb interaction, this piece can be compensated by the
solenoidal magnetic field of the pair. Indeed, this pattern is confirmed with numerical simulations \cite{Loiko:2020htk},
we found linear chains of spinning gauged Q-balls, located
symmetrically with respect to the origin along the symmetry axis. These solutions are classified by the
winding number $n$ and the number of constituents $k$. In Fig.~\ref{Fig6} we displayed an example of such a chain with
6 components. Note that the neighboring complex components of the system are in opposite phases, as expected.
Interestingly, there are two branches of solutions of the system \re{lag-fls}, the chains exist only in a frequency range,
which is restricted from  below  by  some  critical value of angular frequency. The electric repulsion provides a leading
contribution to the interaction energy on the lower in energy branch, in a contrary, the magnetic energy rapidly grows along
the upper, magnetic branch, it extends forward as the frequency increases. Peculiar feature if this branch is that
the strong magnetic field of the vortex destroys the superconductive phase in some region inside the spinning Q-ball
\cite{Loiko:2019gwk,Loiko:2020htk}.

%%%%%%%%%%%%%%%%%%%%%%%%%%%%%%%%%%%%%%%%%%%%%%%%%%%%%%%%%%%%%%%%%%
\section{Monopole-anti Monopole chains}
%%%%%%%%%%%%%%%%%%%%%%%%%%%%%%%%%%%%%%%%%%%%%%%%%%%%%%%%%%%%%%%%%%
As we have seen in previous section, some localized configurations may possess a few
different types of asymptotic fields. In such a situation all of the corresponding
interactions in a pair of separated solitons have to be balanced  to provide a zero net force.
An interesting example of such a system with different types of asymptotic fields is the monopole-antimonopole
chains in the non-Abelian Yang-Mills-Higgs theory
\cite{Rueber,MAP,Kleihaus:2003nj,Kleihaus:2003xz,Kleihaus:2004is,Kunz:2006ex,Shnir:2005te}.

\begin{figure}
\includegraphics[height=4.4cm]{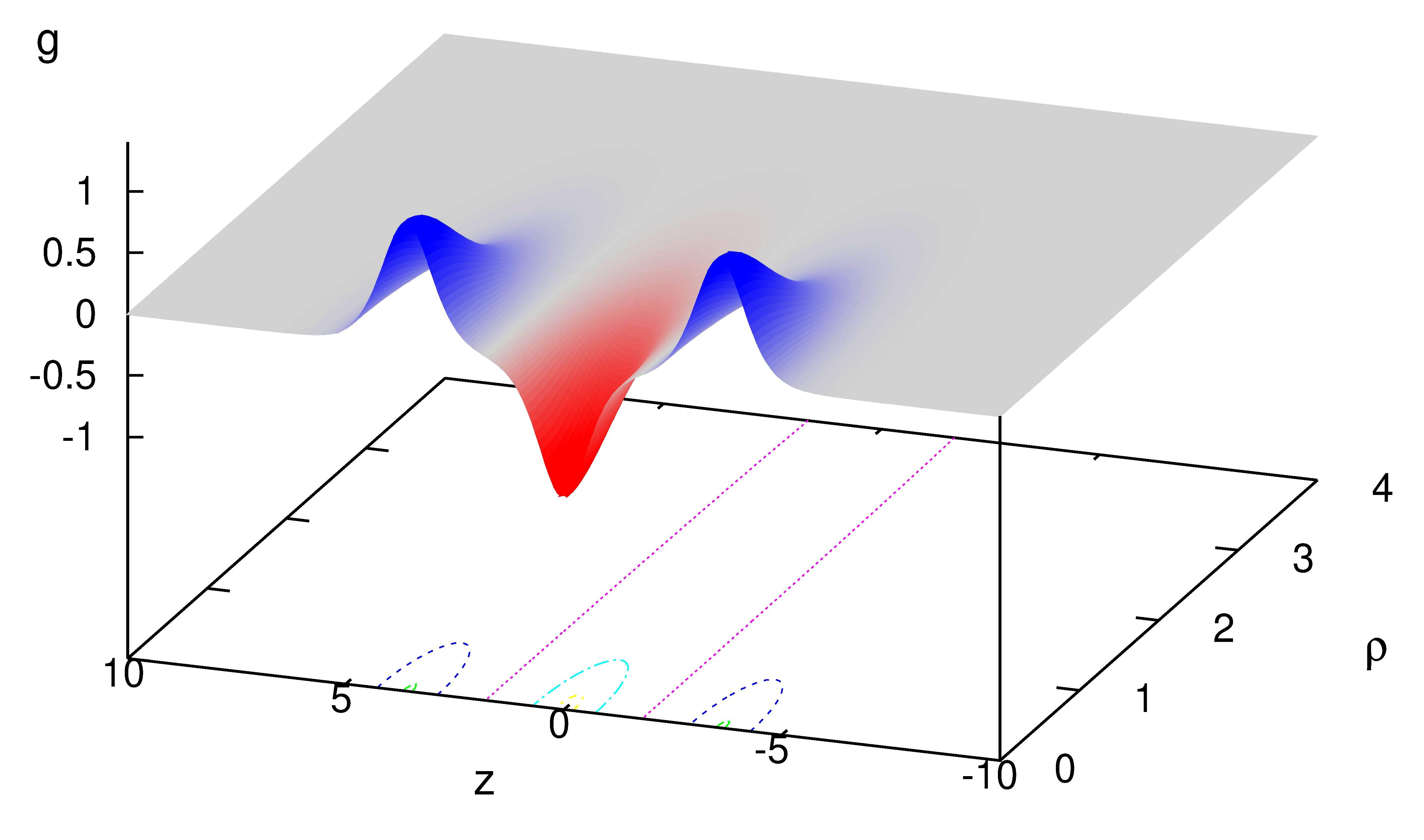}
\includegraphics[height=4.4cm]{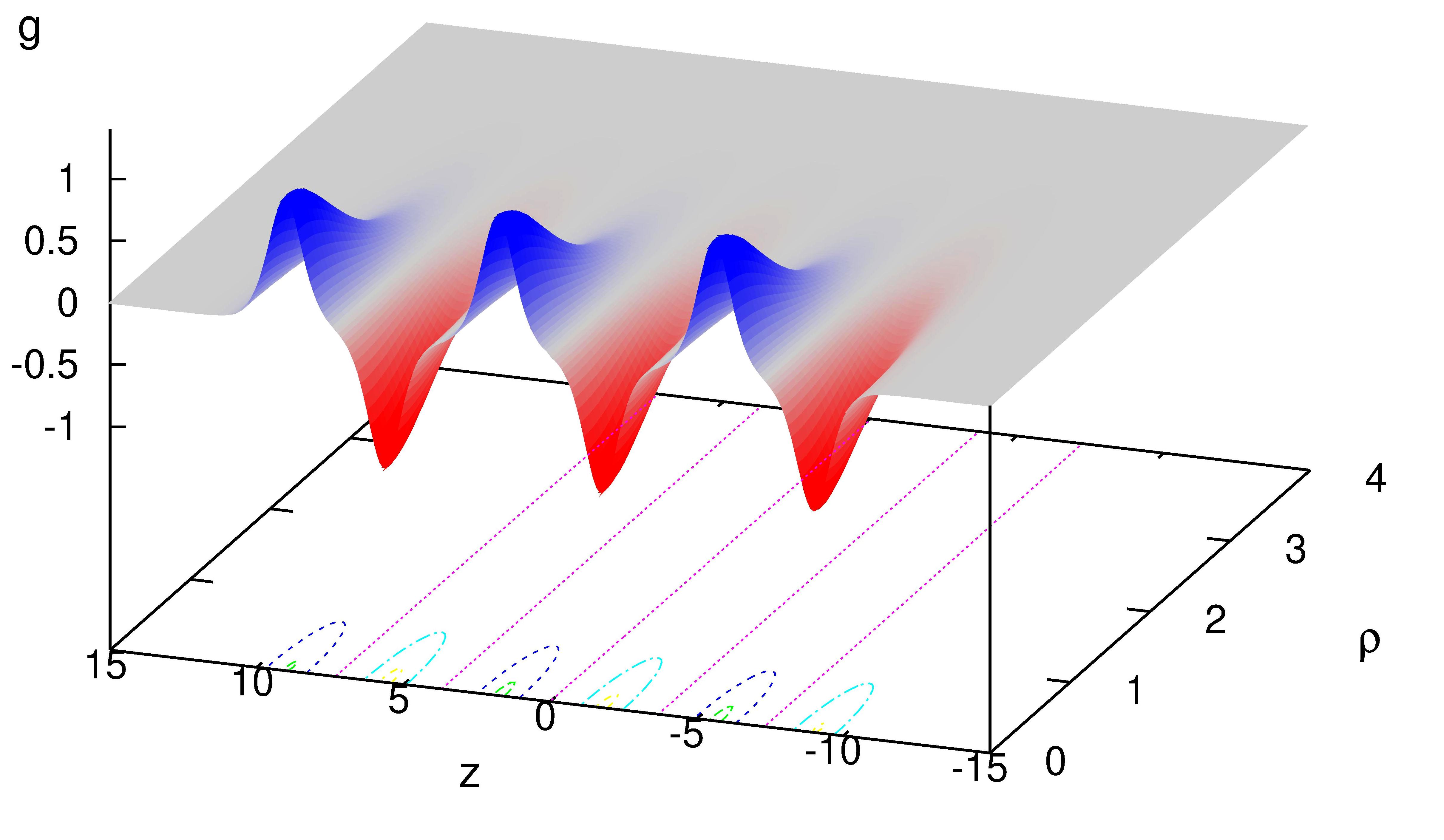}
\includegraphics[height=4.8cm]{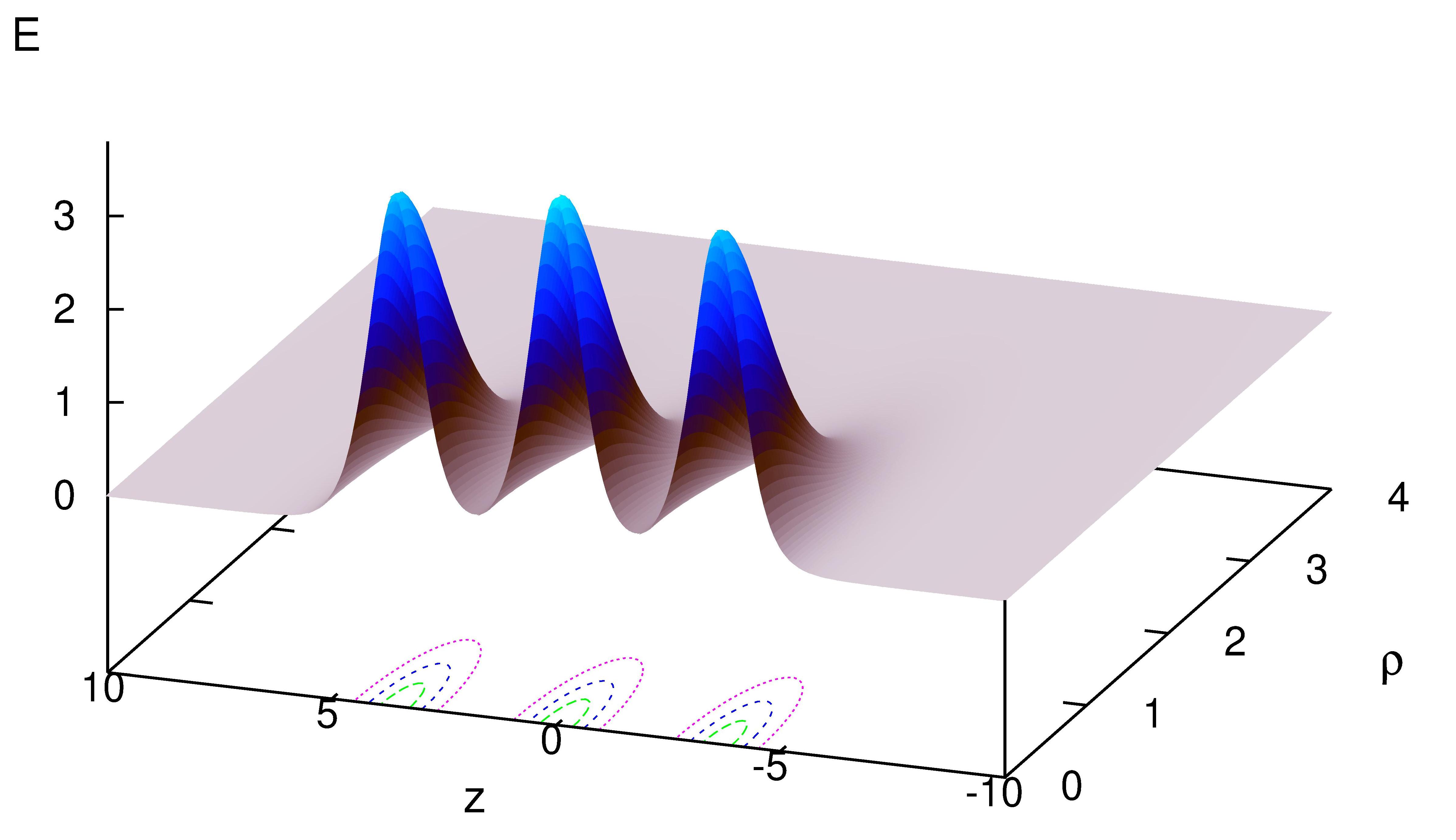}
\includegraphics[height=4.7cm]{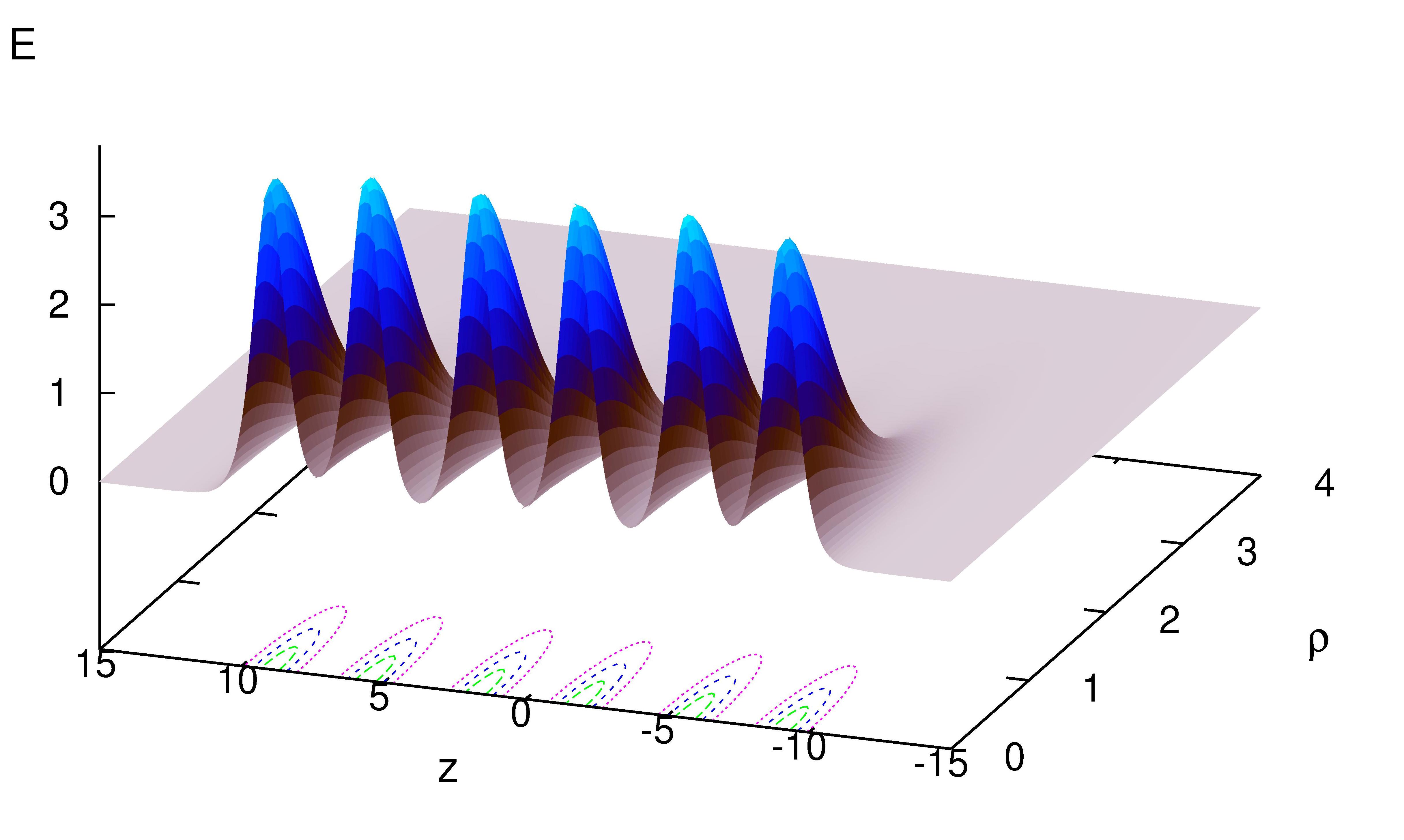}
\includegraphics[height=4.8cm]{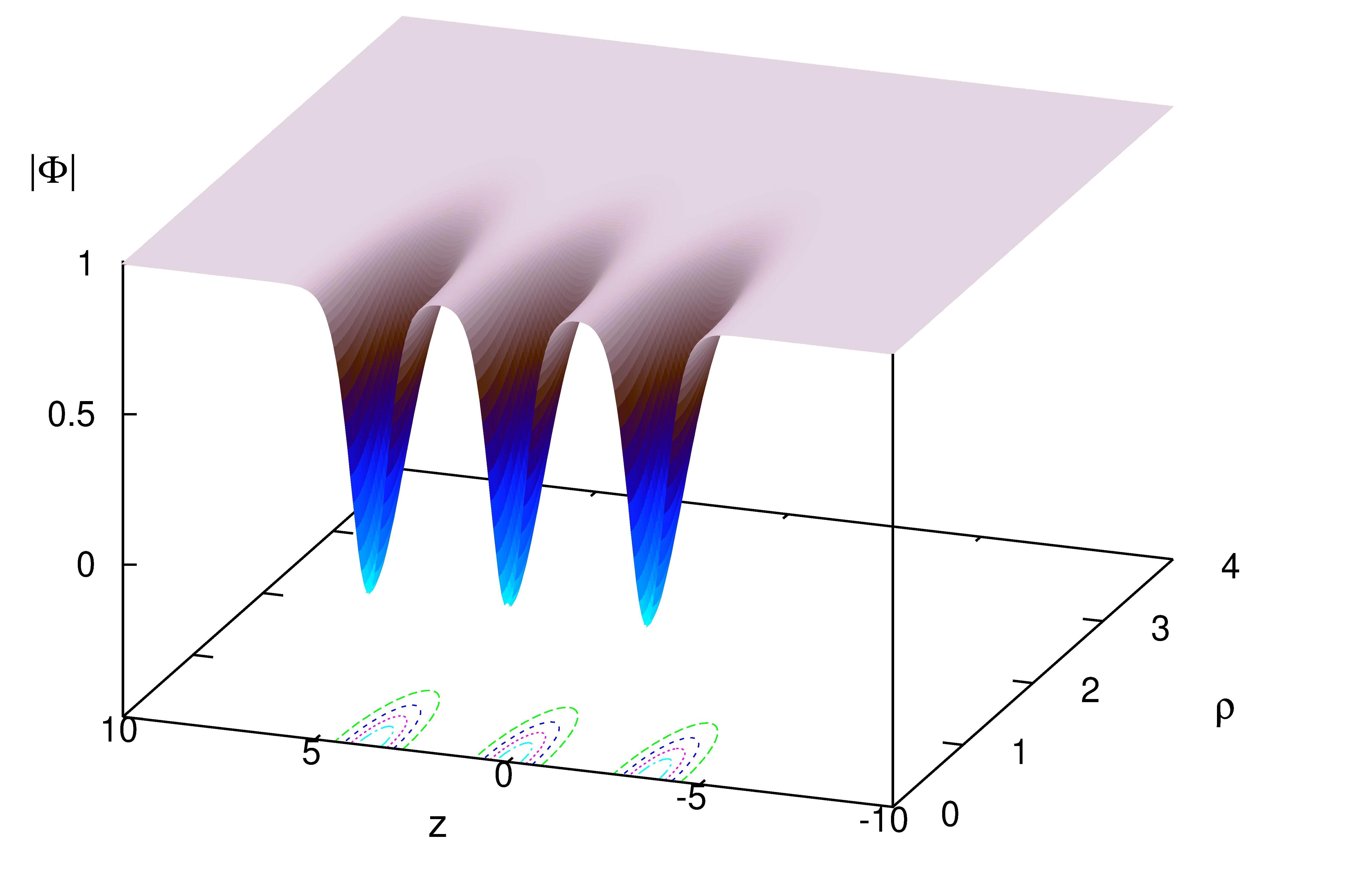}
\includegraphics[height=4.8cm]{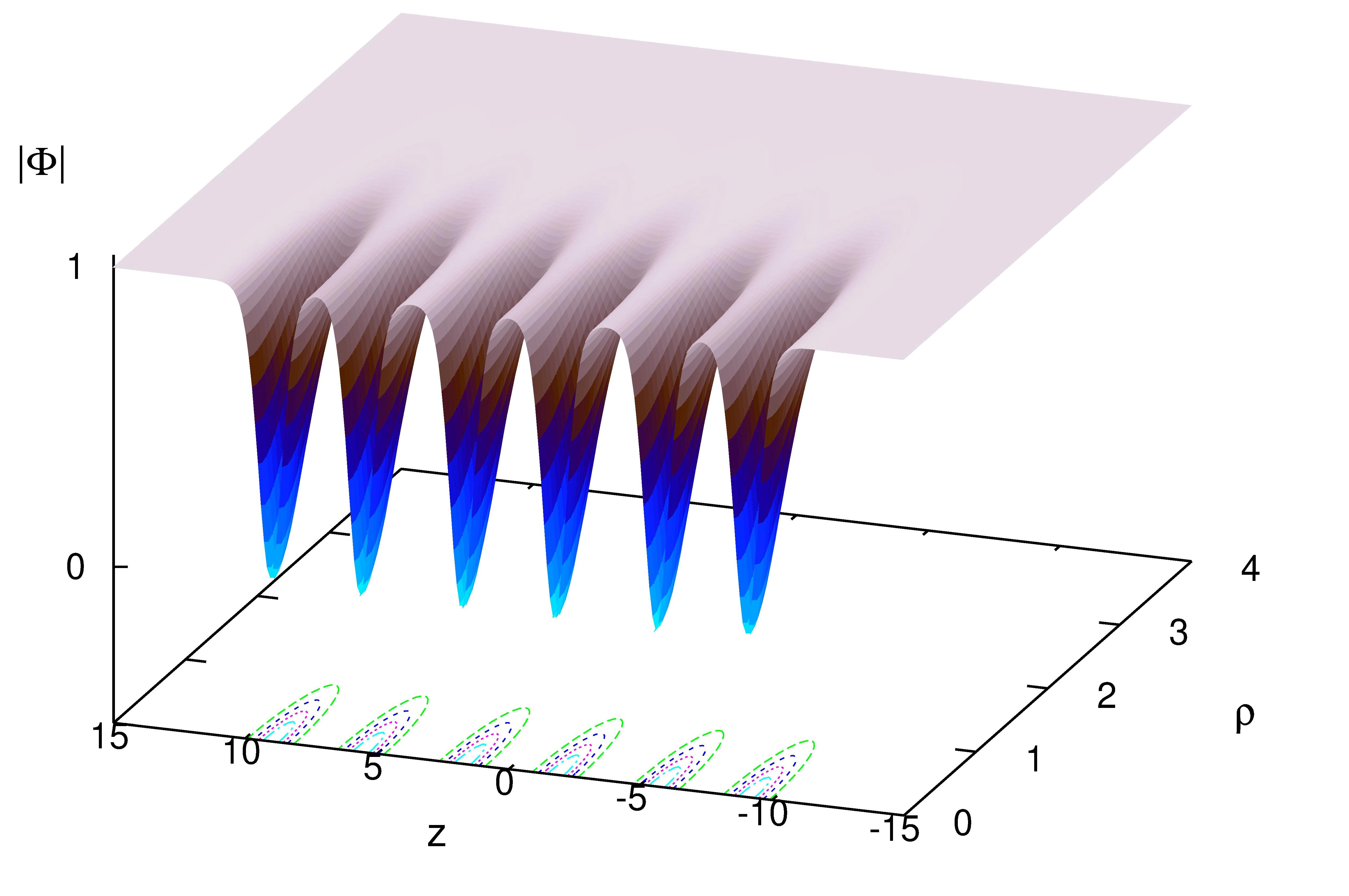}
\caption{\small
Monopole-anti-monopole chains:
The magnetic charge density distributions (upper plots),
the energy density distributions (middle plots))
and the magnitude of the Higgs field  (bottom  plots),
of the $n=1, m=3$ and $n=1, m=6$ chains.}
\label{Fig7}
\end{figure}

The $SU(2)$ Yang-Mills-Higgs theory has the Lagrangian density
\be
L=\frac12~{\rm Tr}~(F_{\mu\nu} F^{\mu\nu}) +\frac{1}{4}~{\rm Tr}~(D_\mu\Phi D^{\mu} \Phi)+ \lambda~{\rm Tr}~(\Phi^2 -1)^2
\label{lag-YMH}
\ee
with gauge potential $A_\mu = A_\mu^a \tau^a$, field strength
tensor $F_{\mu\nu}=\partial_\mu A_\nu-\partial_\nu A_\mu +ie[A_\mu,A_\nu]$, and covariant
derivative of the Higgs field $D_\mu\Phi=\partial_\mu \Phi+ie[A_\mu, \Phi]$. Here $e$ is the gauge coupling and $\lambda$
is the strength of the scalar self-coupling.

The static solutions of the corresponding field
equations can be constructed numerically by employing of
the axially-symmetric ansatz \cite{Kleihaus:2003nj,Kleihaus:2003xz,Kleihaus:2004is}
\begin{eqnarray}
A_\mu dx^\mu &=&
\left( \frac{K_1}{r} dr + (1-K_2)d\theta\right)\frac{\tau_\vphi^{(n)}}{2e}
-n \sin\theta \left( K_3\frac{\tau_r^{(n,m)}}{2e}
                     +(1-K_4)\frac{\tau_\theta^{(n,m)}}{2e}\right) d\vphi\, ,\nonumber \\
\Phi
& = &
H_1\tau_r^{(n,m)}+ H_2\tau_\theta^{(n,m)} \,  ,
\label{ansatzYMH}
\end{eqnarray}
where the $su(2)$ matrices
$\tau_r^{(n,m)}$, $\tau_\theta^{(n,m)}$, and $\tau_\varphi^{(n)}$ are defined as a product
of these vectors with
the usual Pauli matrices $\tau^a$:
\begin{eqnarray}
\tau_r^{(n,m)}  & = &
\sin(m\theta) \tau_\rho^{(n)} + \cos(m\theta) \tau_z \ ,
\nonumber\\
\tau_\theta^{(n,m)} & = &
\cos(m\theta) \tau_\rho^{(n)} - \sin(m\theta) \tau_z \ ,
\nonumber\\
\tau_\varphi^{(n)} & = &
 -\sin(n\varphi) \tau_x + \cos(n\vphi)\tau_y \ ,
\nonumber
\end{eqnarray}
where $\tau_\rho^{(n)} =\cos(n\varphi) \tau_x + \sin(n\varphi)\tau_y $ and $\rho = \sqrt{x^2 + y^2} =
r\sin\theta$. Note that the ansatz (\ref{ansatzYMH}) is axially symmetric, a spatial rotation around the $z$-axis
can be compensated by an Abelian gauge transformation
$U = \exp \{i\omega(r,\theta) \tau_\varphi^{(n)}/2\}$.
Variation of the Lagrangian \re{lag-YMH} yields a
system of six second-order non-linear partial differential equations in the
coordinates $r$ and $\theta$, these equations can be solved numerically, see
\cite{Kleihaus:2003nj,Kleihaus:2003xz,Kleihaus:2004is}.
The well-known
spherically symmetric `t Hooft--Polyakov ansatz is recovered as  we impose the constraints
$K_1=K_3=H_2=0,~ K_2=K_4=K(r),~ H_1 = H(r)$.

The generalized monopoles \re{ansatzYMH} are characterized by two integers, the
winding number $m$ in polar angle $\theta$ and the winding
number $n$ in azimuthal angle $\varphi$. Making use of the usual definition of the topological
charge of the configuration and taking into account the boundary conditions of the profile functions,
we obtain \cite{Kleihaus:2004is}
\be
Q=\frac{1}{8\pi}\int\limits_{S^2}\!{\rm Tr}~(\hat \Phi~ d\hat\Phi\wedge d\hat\Phi)=
\frac{n}{2}\left(1-(-1)^m\right)\, ,
\ee
where $\hat \Phi$ is the $su(2)$ normalized Higgs field. Hence, the configurations
with even values of the winding number $m$ are axially-symmetric
deformations of the topologically trivial sector, while the
configurations with odd values of $m$ are deformations of the fundamental
`t Hooft--Polyakov solution \cite{Kleihaus:2004is}.
Note that here we discuss solutions of the second order equations, they do not satisfy the first order BPS monopole
equations.

Simplest non-trivial solution represent a monopole-anti-monopole pair in a static equilibrium.
The reason of existence of such a solution, a magnetic dipole, is related to an exact
balance of short-range Yukawa interactions mediated be the vector and scalar fields.

Indeed, the pattern of interaction between non-abelian monopoles
does not correspond to a naive picture of electromagnetic Coulomb interaction between two point-like magnetic charges.
First, there is an attractive force between well separated monopoles, it is
mediated by the $A^3_\mu$ component of the Yang-Mills field. However, this field
is massless only on the spatial infinity, such interaction is short-ranged. On the other hand,
there is a scalar attraction mediated by a massive Higgs boson, so the monopoles attract each other with double force.

The situation is different in the BPS limit, then both the gauge and the scalar field possess long-range Coulomb
asymptotic. Further, in such a case repulsive gauge interaction between the monopoles is always balanced by the scalar
interaction for any separation, any system of BPS monopoles can be static.

It was pointed out by Taubes \cite{Taubes} that,
for non-BPS monopoles, the massive vector bosons $A^{\pm}_\mu$
also mediate the short-range Yukawa interactions
between the monopoles and contribute to the interaction energy. Furthermore,
the sign of this contribution to the
net interaction potential depends on the relative orientation of the
monopoles. The monopole-anti-monopole  pair is a  saddle point configuration where the attractive short-range
forces, mediated both by the $A^3_\mu$ vector boson and the
Higgs boson, are balanced by the repulsive interaction due to massive vector bosons $A^{\pm}_\mu$
with opposite orientation in the group space \cite{Shnir:2005te}. The effective net potential of the
interaction between a monopole
and an anti-monopole is attractive for large separation and it is repulsive on a short distance, it resembles as that of
well known Van der Waals molecular potential. The pair is a sphaleron solution in the topologically trivial sector,
it corresponds to the middle of non-contractible loop on the configuration space of the system. This loop corresponds to the
creation of a monopole-antimonopole pair with
relative orientation in the internal space $-\pi$ from the vacuum, separation of the pair, rotation of the monopole by
$2\pi$, and annihilation of the pair back into vacuum \cite{Taubes}.

Furthermore, each topological sector of the Yang-Mills-Higgs model \re{lag-YMH} contains
besides the (multi)monopole solutions further regular, finite
mass solutions, which do not satisfy the first order Bogomolnyi
equations, but only the set of second order field equations,
even for vanishing Higgs potential \cite{Taubes}. Such solutions form saddlepoints of
the energy functional, and possess a mass above the BPS bound.

The simplest solution of that type, $m=2$,$n=1$ monopole-anti-monopole pair, posses two zeros of the Higgs field
located symmetrically on the positive and negative $z$ axis, the peaks of the energy density distribution are
associated with these zeros. Further generalizations of this solution correspond to the chains of monopoles and
anti-monopoles, each carrying charge $n=\pm 1$ in alternating order. In Fig.~\ref{Fig7} we displayed two examples
of these chains with $m=3,6$ and $n=1$ at $\lambda=0.5$. The chains with even number of constituents $m$ are
deformations of the topologically trivial sector, while the chains with odd values of $m$ represent deformations
of the charge $n$ monopole. Positions of the partons in a chain depend on these integers, for $n=1,2$ and
relatively small values of the scalar coupling $\lambda $ the solitons are located on the symmetry axis. Note that
the asymptotic field of the monopole-anti-monopole pair represents a magnetic dipole \cite{Kleihaus:2003nj}
\be
A_\mu dx^\mu \sim \frac{d}{2r}\sin^2 \theta ~\tau_3 d\varphi
\ee
where $d$ is the dipole moment. The emergency of the chains can be explained as formation of the system of aligned dipoles,
just in the same way as Skyrmion chains are formed.

Note that, as the charge $n$ of the monopole and anti-monopole in the chain increases beyond $n = 2$, it becomes
favorable for the monopole-anti-monopole system to form a system of vortex rings,
in which the Higgs field vanishes on a ring centered around the symmetry axis
\cite{Kleihaus:2003xz,Kleihaus:2004is,Kunz:2006ex}. For larger values of the scalar mass
also more complicated configurations can appear, which consist of monopole-anti-monopole pairs or chains as
well as vortex rings, the situation becomes much more complex as gravity is included into
consideration \cite{Kleihaus:2004fh}, and/or an electric charge is added to the monopoles
in the chain \cite{Kleihaus:2005fs}.

Remarkably, that there is certain similarity with the Skyrmion-anti-Skyrmion chains
we discussed previously in the Section 4.
In both cases the soliton-anti-soliton chains represent axially symmetric saddle-point sphaleron-type solutions
which are characterized by two integers,
one of which yields the number of constituents in the chain, and
another corresponds to the absolute value of the topological charge of a component.
Further,  the component of the Skyrme field $\sigma$ shows a clear relation to
the corresponding behavior of the magnitude of the Higgs field $|\Phi|$ in the Yang-Mills-Higgs system \cite{Krusch:2004uf}.
However, the dipole-dipole interaction
between the components of the SAS pair is weaker than the short-range Yukawa interactions in the monopole-anti-monopole pair,
in the flat space the SAS pair may exist only if the topological charge of the components is higher than two \cite{Krusch:2004uf}.
However, the effect of the pion mass term \cite{Shnir:2009ct} or coupling to gravity \cite{Shnir:2015aba,Shnir:2020hau},
open a possibility for existence of the Skyrmion-anti-Skyrmion chains with constituents carrying unit topological charge.

Our final remark is related with a possibility to construct Euclidean counterparts
of the MAP solutions in four dimensional Yang-Mills theory \cite{Radu:2006gg}.
These non-self-dual instanton-antiinstanton saddle point configurations are obtained numerically by analogy with the
construction of the above-mentioned  axially-symmetric MAP solution. Since the holonomy of
Yang-Mills instantons provides a very good approximations to Skyrmions \cite{Atiyah:1989dq}, it is of no surprise that
the holonomy  of  the  chains  of  interpolating calorons-anticalorons gives  a
nice  approximation  to  the  corresponding Skyrmion-anti-Skyrmion chains \cite{Shnir:2013ova}.

%%%%%%%%%%%%%%%%%%%%%%%%%%%%%%%%%%%%%%%%%%%%%%%%%%%%%%%%%%%%%%%%%%%%%%%%%%%%%%
\section{Conclusions}
%%%%%%%%%%%%%%%%%%%%%%%%%%%%%%%%%%%%%%%%%%%%%%%%%%%%%%%%%%%%%%%%%%%%%%%%%%%%%%
The main purpose of this short review was to provide a comparative analysis of different types of linear
chains of non-self dual solitons in various models. Such solutions exist because of balancing of repulsive
and attractive interactions, in most cases they represent sphaleron-like field configurations.
Apparently, any multisoliton solution in one spatial dimension represents a chain, however such static configuration may exist
only if there is a dynamic equilibrium between the repulsive and attractive forces. A particular example is a real scalar
theory coupled to the Dirac fermions, here scalar repulsion between the kinks is evened out by fermionic exchange interaction.
Other higher-dimensional examples include
self-gravitating chains of boson stars, where the scalar repulsion is compensated by the gravitational attraction of
the solitons and Skyrmion-anti-Skyrmion chains, which are stabilized by the dipole-dipole interactions. The pattern of
interactions between the constituents of linear chains of spinning gauged Q-balls in two-component
Friedberg-Lee-Sirlin model is more complicated, it includes both scalar and electromagnetic forces. Similarly, the
monopole-anti-monopole chains exist because of precise balance of the scalar and gauge short-range Yukawa
interactions mediated by the Higgs boson and by the massive vector bosons $A^{\pm}_\mu, A^3_\mu$, respectively.
Notably, there is a certain
similarity between the monopole-anti-monopole chains in the non-abelian Yang-Mills-Higgs model, and
Skyrmion-anti-Skyrmion chains, related with the Atiyah-Manton construction \cite{Atiyah:1989dq}.

%%%%%%%%%%%%%%%%%%%%%%%%%%%%%%%%%%%%%%%%%
\section*{Acknowledgements}
%%%%%%%%%%%%%%%%%%%%%%%%%%%%%%%%%%%%%%%%%

I am grateful to Carlos Herdeiro, Burhard Kleihaus, Jutta Kunz, Viktor Loiko, Ilya Perapechka, Eugen Radu, Albert Samoilenka,
Tigran Tchrakian and Gleb Zilin
for valuable collaboration, many results of our joint work are reviewed in this brief survey. I would like to acknowledge
discussions with Yuki Amari, David Foster, Muneto Nitta, Derek Harland, Nick Manton, Tomasz Roma\'nczukiewicz, Ivan Smalyukh,
Paul Sutcliffe, Nobuyuki Sawado and Wojtek Zakrzewski.

The work was supported by Ministry of Science and High Education of Russian Federation, project FEWF-2020-0003.
Computations were performed on the cluster HybriLIT (Dubna).

 %%%%%%%%%%%%%%%%%%%%%%%%%%%%%%%%%%%%%%%%%%%%%%%%%%%%%%%%%%%%%%%%%%
 \begin{small}
 
%%%%%%%%%%%%%%%%%%%%%%%%%%%%%%%%%%%%%%%%%%%%%%%%%%%%%%%%%%%%%%%%%%%%%%%%%%%%%%
 \end{small}

 \end{document}